\documentclass{article}
\usepackage{graphicx} 
\usepackage[table]{xcolor}
\usepackage{color, colortbl}

\usepackage{authblk}
\providecommand{\keywords}[1]{\textbf{\textit{Keywords:}} #1}

\usepackage{amsmath}
\usepackage{enumitem}

\title{Preparing for the Post-Quantum Era: Quantum-Ready Architecture for Security and Risk Management (QUASAR) – A Strategic Framework for Cybersecurity}
\author[1]{Abraham Itzhak Weinberg}
\affil[1]{AI-WEINBERG, AI Experts, Tel Aviv, Israel, aviw2010@gmail.com}

\begin{document}
\maketitle
\begin{abstract}
As quantum computing progresses, traditional cryptographic systems face the threat of obsolescence due to the capabilities of quantum algorithms. This paper introduces the Quantum-Ready Architecture for Security and Risk Management (QUASAR), a novel framework designed to help organizations prepare for the post-quantum era. QUASAR provides a structured approach to transition from current cryptographic systems to quantum-resistant alternatives, emphasizing technical, security, and operational readiness. The framework integrates a set of actionable components, a timeline for phased implementation, and continuous optimization strategies to ensure ongoing preparedness. Through performance indicators, readiness scores, and optimization functions, QUASAR enables organizations to assess their current state, identify gaps, and execute targeted actions to mitigate risks posed by quantum computing. By offering a comprehensive, adaptable, and quantifiable strategy, QUASAR equips organizations with the tools necessary to future-proof their operations and secure sensitive data against the impending rise of quantum technologies.
\end{abstract}


\section{Introduction}

As quantum computing rapidly advances, the potential for its disruptive impact on current cryptographic systems becomes increasingly evident \cite{mavroeidis2018impact}. Traditional encryption methods, which form the backbone of data security across industries, are vulnerable to quantum algorithms capable of efficiently solving problems that are computationally infeasible for classical computers \cite{ajala2024exploring}. To mitigate the risks posed by this technological shift, organizations must adopt a proactive approach to prepare for the post-quantum era\footnote{Post quantum refers to the creation of cryptographic algorithms believed to be secure against attacks by quantum computers.}. This paper outlines key principles for organizations to follow in transitioning to post-quantum, quantum-resistant cryptographic solutions, ensuring the continued confidentiality, integrity, and authenticity of sensitive data. By focusing on critical areas such as risk assessment, cryptographic infrastructure analysis, technical preparation, and organizational strategy, this guide aims to equip organizations with the necessary tools and knowledge to navigate the challenges and opportunities presented by the advent of quantum computing  \cite{baseri2024cybersecurity,geremew2024preparing}.\\
To support this transition, we introduce a novel framework: Quantum-Ready Architecture for Security and Risk Management (QUASAR). QUASAR provides a structured, strategic approach for organizations to assess, plan, and implement quantum-resilient cybersecurity measures. It goes beyond theoretical principles by offering actionable guidance tailored to the evolving threat landscape. The framework integrates key components such as risk management, cryptographic agility, organizational alignment, and continuous adaptation to help businesses stay ahead of quantum-era threats. Through QUASAR, we aim to bridge the gap between awareness and action—empowering organizations to future-proof their security posture in a practical, scalable, and resilient manner.\\

The uniqueness of this paper lies in its introduction of the QUASAR framework, which offers a comprehensive, actionable, and quantifiable approach to post-quantum cybersecurity readiness. This work focuses not only on cryptographic algorithm development and theoretical risk assessments, but also bridges the gap between technical innovation and organizational strategy. Moreover, QUASAR integrates component-based readiness matrices, implementation phases, performance optimization models, and success metrics into a unified, adaptable framework tailored for real-world deployment. Its holistic design addresses technical, operational, and strategic dimensions, enabling organizations not only to transition to quantum-resistant systems but to do so in a structured, measurable, and sustainable way. This multidimensional perspective, grounded in both mathematical rigor and practical application, sets the framework apart as a pioneering contribution to the field of quantum-era cybersecurity preparedness.\\

The paper is organized as follows. It begins with the Introduction, which outlines the emerging threat landscape posed by quantum computing and highlights the urgency for organizations to transition toward quantum-resistant systems. The next section, Post-Quantum Approaches and Algorithms for Cybersecurity, reviews the current state of quantum-safe cryptographic solutions and their implications for modern security architectures. The paper then introduces a high-level Framework and Pipeline for organizational transformation in the post-quantum context, setting the stage for the core proposal. \\ Central to the paper is the Quantum-Ready Architecture for Security and Risk Management (QUASAR) Model Framework, which presents a detailed, component-based structure for guiding organizations through the readiness process. The Implementation Framework and Implementation Phases sections translate this model into actionable steps and a phased deployment strategy. To support ongoing evaluation and alignment, the Success Metrics and Timeline Framework sections offer quantifiable indicators and scheduling tools. The paper further emphasizes adaptability through the Continuous Improvement section, ensuring long-term resilience. Finally, the Discussion consolidates the findings, reflects on practical considerations, and explores future research directions, reinforcing the relevance and scalability of the QUASAR framework in addressing quantum-era cybersecurity challenges

\subsection{Risk Assessment and Inventory}
Post-quantum preparedness begins with a thorough risk assessment and inventory of an organization's cryptographic systems and sensitive assets \cite{geremew2024preparing}. The first step involves identifying and documenting all systems that rely on cryptographic protocols, ensuring that every application and infrastructure component is accounted for \cite{aydeger2024towards}. It is equally important to catalog sensitive data that must remain secure for extended periods, as this data will require long-term protection against quantum threats \cite{hasan2024framework}. \\
A comprehensive mapping of data flows and cryptographic dependencies within the organization is essential to understand how data moves and interacts with cryptographic systems \cite{nather2024migrating}. This mapping helps prioritize systems based on their vulnerability to quantum computing threats, allowing organizations to address the most critical systems first \cite{joseph2022transitioning}. \\
Alongside asset identification, a detailed cryptographic infrastructure analysis must be conducted. This includes cataloging all cryptographic protocols in use, documenting key lengths and algorithms for each application, and identifying those systems that rely on potentially vulnerable algorithms such as Rivest–Shamir–Adleman (RSA), Elliptic-curve cryptography (ECC), or Diffie Hellman (DH) key exchange \cite{fathalla2024beyond}. Additionally, mapping dependencies within the Public Key Infrastructure (PKI) and certificates is crucial to understanding the broader cryptographic landscape, ensuring that vulnerabilities are detected and mitigated before quantum computing becomes a practical threat.

\subsection{Technical Preparation}
Technical preparation for Post Quantum Cryptography (PQC) involves ensuring that systems are capable of adapting to new cryptographic algorithms as quantum computing advances \cite{geremew2024preparing}. One critical aspect of this is implementing crypto-agility, which entails designing systems with the flexibility to swap cryptographic algorithms as needed \cite{grote2019review}. This can be achieved through modular cryptographic interfaces and the creation of abstraction layers for cryptographic operations, allowing for easy updates and the rapid replacement of algorithms \cite{joseph2022transitioning}.\\ 
Developing procedures for swift algorithm replacement ensures that organizations can quickly respond to emerging quantum threats without significant disruption to their systems \cite{bishwas2024strategic}. In parallel, organizations must closely monitor the National Institute of Standards and Technology (NIST) PQC standardization process to stay informed about the progress and finalization of quantum-resistant algorithms \cite{kappler2022post,joshi2024guarding}. Once post-quantum algorithms are approved, they should be tested in non-production environments to assess their compatibility, performance, and potential impact on system operations. Evaluating the performance of these algorithms and understanding their implementation requirements across different systems is crucial to ensuring seamless integration into existing infrastructures while maintaining security and efficiency in a post-quantum era \cite{demir2025performance}.

\subsection{Organizational Strategy}
In effective organizational strategy for PQC involves both the development of comprehensive policies and strategic resource planning to ensure a smooth and secure transition. Policy development should focus on creating clear standards for quantum-safe cryptographic practices, defining guidelines for the implementation of new systems that meet post-quantum security requirements \cite{kong2024realizing}.\\
Additionally, organizations must establish transition timelines for legacy systems, ensuring a structured approach to upgrading or replacing outdated cryptographic methods. Procurement policies should also be updated to prioritize the acquisition of quantum-safe technologies, ensuring that all future systems and software are compatible with post-quantum standards. In parallel, resource planning is crucial to support the transition. This includes allocating sufficient budget for post-quantum transitions, identifying the necessary expertise and training to upskill staff in quantum-safe technologies, and planning for increased computational and storage requirements due to the larger cryptographic keys and data \cite{bishwas2024strategic}. \\
Furthermore, establishing partnerships with quantum security experts is essential for gaining insights into emerging threats and ensuring that the organization remains at the forefront of post-quantum security developments. This holistic approach enables organizations to effectively prepare for the post-quantum era while maintaining operational security and efficiency.

\subsection{Implementation Roadmap}
The implementation of a Post Quantum Cryptographic (PQC) strategy requires a phased approach, with actions tailored to near-term, medium-term, and long-term objectives to ensure a smooth and secure transition. In the near term (1-2 years), organizations should initiate a comprehensive cryptographic inventory, cataloging all systems and protocols in use. This phase should also focus on implementing crypto-agility in new systems, allowing them to seamlessly integrate post-quantum algorithms as they become available \cite{marchesi2025survey}. \\
Security teams must be trained to understand the potential threats posed by quantum computing, and testing of NIST-approved post-quantum algorithms should begin in non-production environments to evaluate their compatibility and performance \cite{okika2025assessing}. Over the medium term (2 to 5 years), hybrid solutions that combine classical and post-quantum algorithms should be deployed to protect systems during the transition period \cite{ricci2024hybrid}. \\ Procurement processes should be updated to prioritize quantum-safe technologies, while high-priority systems are migrated to quantum-resistant solutions. Enhanced monitoring capabilities must also be implemented to detect potential vulnerabilities and ensure the effectiveness of security measures. In the long term (5+ years), organizations should complete the migration to quantum-resistant algorithms and decommission systems that are still vulnerable to quantum threats. Ongoing awareness of developments in quantum computing is essential, along with regular reviews and updates to the organization's quantum security posture to ensure continuous protection as the landscape evolves. This roadmap allows organizations to gradually adapt to PQC while maintaining robust security at each stage of the transition.

\subsection{Business Continuity}
Business continuity in the context of PQC necessitates proactive management of both partner relationships and customer interactions to ensure seamless transitions and sustained security. Critical partners, especially those integral to the organization's supply chain, must be assessed for their quantum readiness, with quantum security requirements included in contracts to mitigate risks associated with the adoption of new technologies \cite{geremew2024preparing}.\\
Partner transition timelines should be developed to align with the organization’s own migration strategy, ensuring synchronization across the supply chain. Communication protocols for security updates should also be established, facilitating quick and clear dissemination of relevant information regarding quantum security advancements. \\
On the customer side, identifying customer-facing systems that require updates is essential to ensure that users are not impacted by the transition to quantum-resistant solutions. A well-defined customer communication strategy should be developed, outlining the steps being taken to maintain service continuity during the transition \cite{aydeger2024towards}. Additionally, planning for service transitions and maintaining backward compatibility where necessary will help minimize disruption, providing customers with confidence in the organization’s ability to safeguard their data against quantum threats while ensuring a smooth, uninterrupted experience \cite{dasquantum}.

\subsection{Compliance and Governance}
Compliance and governance play a critical role in ensuring that organizations remain aligned with regulatory requirements and manage quantum security risks effectively. To prepare for regulatory changes, it is essential to continuously monitor emerging quantum security regulations and document the specific compliance requirements that apply to the organization. Establishing robust audit procedures will help ensure adherence to these regulations, while maintaining evidence of quantum-safe transitions is vital for demonstrating compliance during regulatory reviews \cite{csenkey2023post}. \\
In parallel, risk management frameworks must be updated to incorporate the unique threats posed by quantum computing \cite{csenkey2023post}. This involves integrating quantum risks into existing security models and developing targeted mitigation strategies to address potential vulnerabilities. Regular reviews of the organization’s quantum security posture are necessary to adapt to evolving threats, ensuring that both compliance and risk management efforts remain proactive and effective in safeguarding against the future impacts of quantum technologies.

\subsection{Best Practices}
Implementing best practices in PQC requires a comprehensive approach that combines robust security principles and meticulous documentation. One foundational security principle is maintaining a defense-in-depth strategy, which ensures multiple layers of protection against potential quantum threats \cite{rayhan2024cybersecurity}. Complementing this, the adoption of a zero-trust architecture—where trust is never implicitly granted, and continuous verification is required—further strengthens security across all access points \cite{joshi2024emerging}.\\
Regular security assessments and continuous monitoring and logging are also critical to identifying vulnerabilities and ensuring real-time detection of potential security breaches \cite{hasan2024framework}. In parallel, thorough documentation is essential for maintaining an organized and auditable cryptographic environment. This includes detailed transition plans outlining the migration strategy to quantum-safe solutions, as well as documentation of algorithm choices and the rationale behind them. It is equally important to keep accurate records of all cryptographic assets, ensuring that every key and protocol is accounted for. Regular reviews and updates of this documentation are necessary to reflect changes in both the cryptographic landscape and the organization’s evolving security posture, ensuring preparedness for future threats while maintaining compliance with quantum-safe standards.

\subsection{Training and Awareness}
Training and awareness are crucial components in preparing an organization for the PQC transition. For technical staff, foundational knowledge of quantum computing is essential to understand the potential threats it poses and the need for quantum-resistant solutions \cite{dvorak2024leveraging}. In addition to this, training in the implementation of PQC ensures that staff are well-equipped to deploy, manage, and troubleshoot new algorithms \cite{marmebro2024investigation}. \\
Security testing and validation training is also important to ensure that the post-quantum systems function as expected under real-world conditions, and that incident response procedures are in place to swiftly address any vulnerabilities or attacks \cite{joshi2024guarding}. On the non-technical side, general awareness programs should include executive briefings on the potential impacts of quantum computing, ensuring that leadership understands the strategic importance of transitioning to quantum-safe cryptography \cite{hughes2022assessing}. Department-specific impact assessments help various teams recognize how quantum threats may affect their specific functions. Regular updates on the transition's progress, combined with clear communication of role-specific responsibilities, ensure that all employees, regardless of their role, are aligned with the organization’s overall security strategy and are prepared to contribute to a smooth and secure transition to a post-quantum world.

\section{Post Quantum Approaches and Algorithms for Cybersecurity}

As the advent of quantum computing looms closer, traditional cryptographic methods that rely on the hardness of problems like integer factorization and discrete logarithms are increasingly vulnerable to quantum attacks \cite{meshram2015efficient}. In response, PQC has emerged as a critical field of research and development, focusing on creating algorithms that can withstand the capabilities of quantum computers.\\
This section provides an overview of the leading PQC approaches and algorithms designed to secure data against the potential threats posed by quantum computing. From lattice-based cryptography to hash-based signatures and code-based encryption, these approaches offer promising solutions for ensuring the confidentiality, integrity, and authenticity of information in a quantum-enabled future \cite{guneysu2012practical}. As quantum-resistant standards are being developed and refined, understanding these algorithms and their practical implementations is essential for preparing systems to operate securely in the post-quantum era.\\
One of the traditional premises for encryption is One Way  Function (OWF)\footnote{A function that can be efficiently computed for any input, but is difficult to reverse when given the output corresponding to a random input}. In 1995, Russell Impagliazzo introduced five complexity ``worlds'' \cite{impagliazzo1995personal}:
\begin{enumerate}
    \item \textbf{Algorithmica}: \( P = NP \), with all the remarkable implications that follow from this.
    \item \textbf{Heuristica}: \( NP \)-complete problems are hard in the worst case (\( P \neq NP \)), but can be efficiently solved on average.
    \item \textbf{Pessiland}: Average-case \( NP \)-complete problems exist, but one-way functions do not. This means that generating hard instances of \( NP \)-complete problems with known solutions is not possible.
    \item \textbf{Minicrypt}: One-way functions exist, but public-key cryptography is impossible.
    \item \textbf{Cryptomania}: Public-key cryptosystems exist, enabling secure communication.
\end{enumerate}
Most researchers still believe we are in ``Cryptomania,'' and the evidence supporting this view has remained largely unchanged, with little progress made in collapsing the different complexity worlds.\\
The developments is Shor's algorithm \cite{ugwuishiwu2020overview}, which demonstrates that if we replace \( P \) with \( BQP \footnote{Bounded-error Quantum Polynomial time - A complexity class of decision problems solvable by a quantum computer in polynomial time with a small error margin.} \), many widely used public-key cryptosystems would be insecure. However, due to the existence of lattice-based cryptosystems, the prevailing assumption is that we are still in ``Cryptomania,'' even in this scenario—though the consensus on this may be slightly weaker than in the classical case. In fact, even in the classical case, there seems to be far more evidence supporting the existence of one-way functions (``Minicrypt'') than for the existence of public-key encryption (``Cryptomania''). Nevertheless, the considerable efforts spent attempting to break various public-key cryptosystems provide substantial evidence for the latter as well.\\
The conclusion from the above discussion is that the computational hardness required for quantum cryptography cannot rely solely on OWF. We refer to works that demonstrate quantum commitments and one-time secure digital signatures can exist without the need for quantum-secure classical cryptographic primitives, such as OWF, by showing that pseudorandom quantum states are sufficient for their construction \cite{morimae2022quantum}.\\
Additional works also explore different approaches to cryptography based on chaos theory \cite{weinberg2025dynamic}. For example, the Dynamic Data Defense in Motion Chaos Encryption (DaChE) algorithm can replace traditional methods and help avoid situations where data is harvested and encrypted once post-quantum cryptography becomes mature.

\subsection{Lattice-Based Cryptography}
Lattice-based cryptography represents one of the most promising PQC approaches, offering robust security in the face of quantum computing threats. A key algorithm within this category is CRYSTALS-Kyber, a key encapsulation mechanism based on the Module Learning With Errors (MLWE) problem. Chosen by the National Institute of Standards and Technology (NIST) as the primary standard for post-quantum key encapsulation, CRYSTALS-Kyber is notable for its excellent performance and scalability across different security levels, including Kyber512, Kyber768, and Kyber1024 \cite{bos2018crystals}.\\
These variations provide organizations with flexible options, allowing them to choose the appropriate security strength based on their specific needs while balancing key size and computational efficiency. The algorithm's resilience to quantum attacks, combined with its efficient implementation, makes it a strong candidate for future-proofing cryptographic systems against the advent of quantum computing.\\
CRYSTALS-Dilithium is a PQC algorithm designed for digital signatures, based on the MLWE problem, which offers strong resistance against quantum computing threats. Selected by NIST for standardization as a post-quantum digital signature scheme, CRYSTALS-Dilithium is distinguished by its strong security foundations rooted in hard lattice problems\footnote{Lattice problems are optimization challenges based on mathematical lattices, whose assumed hardness underpins secure lattice-based cryptography. Some are NP-hard and average-case hard, making them useful for testing cryptographic security, while worst-case hard problems can form the basis of highly secure schemes \cite{ajtai1996generating}.}, making it highly resilient to quantum attacks \cite{lyubashevsky2020crystals}. The algorithm provides robust security while maintaining moderate signature sizes, a key advantage for practical deployment. Additionally, it offers multiple variants, each tailored to different security levels, allowing organizations to choose the optimal balance between security and performance. CRYSTALS-Dilithium is also recognized for its efficient implementation possibilities, with relatively compact key sizes and excellent performance characteristics, making it an attractive solution for integrating post-quantum digital signatures into future cryptographic systems.

\subsection{Hash-Based Signatures}
Hash-based signatures, such as SPHINCS+ and eXtended Merkle Signature Scheme (XMSS), provide quantum-resistant alternatives to traditional digital signature schemes, leveraging the security of hash functions rather than relying on lattice or number-theoretic assumptions \cite{mohan2023hash}. SPHINCS+ is a stateless hash-based signature scheme, selected by NIST as an alternate signature algorithm, that offers a strong security foundation based solely on the security of hash functions. The standard uses the Sphincs+ algorithm, now renamed StateLess Hash-based Digital Signature Algorithm (SLH-DSA) \cite{saarinen2024accelerating}.\\ 
While it produces larger signatures compared to other algorithms, its robustness against quantum threats makes it a compelling choice for long-term data integrity. On the other hand, XMSS is a stateful hash-based signature scheme, standardized in RFC 8391\footnote{The Request for Comments (RFC) 8391 publication, issued by key Internet standards bodies, describes the XMSS.}, that is highly efficient in specific, limited-use cases \cite{hulsing2018rfc}.\\ However, it requires careful state management to ensure security and prevent signature reuse. Both algorithms are based on well-understood hash function security, providing quantum resistance even against future quantum algorithms, making them conservative choices with strong security guarantees. These hash-based signatures are particularly suitable for use cases where state management is feasible and where strong, long-term security is paramount.

\subsection{Multivariate Cryptography}
Multivariate cryptography\footnote{Multivariate cryptography uses multivariate polynomials over finite fields for asymmetric encryption. When the polynomials are quadratic, the system is called multivariate quadratic, and solving these equations is NP-complete.}, particularly through algorithms like Rainbow, offers a promising approach to PQC by leveraging the difficulty of solving systems of multivariate quadratic equations \cite{dey2023progress} . Rainbow, a multivariate signature scheme, is known for its compact signature sizes and its ability to generate and verify signatures quickly, making it highly efficient in environments requiring fast cryptographic operations. \\
However, it is important to note that the original version of Rainbow was found to be vulnerable, although research continues to refine and improve the scheme \cite{dey2023progress}. Despite its challenges, multivariate cryptography remains an active area of research due to its potential for small signature sizes and fast signature generation, which are desirable features for resource-constrained systems. The security of multivariate cryptographic schemes, however, remains complex and is the subject of ongoing analysis, as researchers continue to explore ways to enhance their robustness and address existing vulnerabilities, ensuring their viability in a post-quantum world.

\subsection{Code-Based Cryptography}
Code-based cryptography, exemplified by algorithms such as Classic McEliece, provides a robust foundation for PQC by relying on the hardness of error-correcting codes \cite{balamurugan2021post}. Classic McEliece, selected by NIST as an alternate key encapsulation mechanism, is notable for its strong security guarantees, underpinned by decades of resistance to cryptanalysis \cite{kuznetsov2021performance}. Although it requires very large public keys, this tradeoff is compensated by fast encryption and decryption operations, making it suitable for applications where computational speed is critical despite the large key sizes.\\
The algorithm is based on well-studied problems from coding theory, which have been extensively analyzed for security over the years, providing confidence in its resilience against quantum attacks. Despite its large key sizes, the fast operational efficiency and established security make code-based cryptography, particularly Classic McEliece, a compelling option for specific applications requiring long-term quantum resistance.

\subsection{Supersingular Isogeny-Based Cryptography}
Supersingular isogeny-based cryptography, particularly exemplified by algorithms like Supersingular Isogeny Key Encapsulation (SIKE), leverages the mathematical complexity of isogenies between supersingular elliptic curves to provide quantum-resistant security \cite{stratil2021supersingular}. SIKE offers the advantage of the smallest key sizes among PQC schemes, making it highly attractive for applications where compact keys are essential \cite{costello2021case}. \\
However, the algorithm's complex mathematical foundation and the challenges surrounding its implementation have led to ongoing research aimed at improving its security and efficiency. The original version of SIKE was found to be vulnerable, but it remains an active area of development as researchers work on refining and enhancing its robustness \cite{mishra2025survey}. Despite these challenges, supersingular isogeny-based cryptography holds promise due to its unique properties, particularly the small key sizes, and continues to be a focus of research for new and improved variants that may offer stronger security and practical deployment in a post-quantum landscape \cite{stratil2021supersingular}.

\subsection{Hybrid Approaches}
Hybrid cryptographic approaches combine classical and post-quantum algorithms or multiple post-quantum algorithms to provide enhanced security during the transition to quantum-resistant systems. The first strategy, using both classical and post-quantum algorithms together, offers a security hedge by maintaining the reliability of traditional cryptography, such as RSA or ECC, while integrating post-quantum algorithms like Kyber\footnote {Kyber and Dilithium are PQC schemes with distinct datapath architectures. Dilithium uses a 23-bit prime modulus, requiring 23-bit arithmetic units, whereas Kyber uses a 12-bit modulus, needing only 12-bit adder, subtractor, and multiplier units.} or Dilithium \cite{ricci2024hybrid} . This combination ensures resilience against both current and emerging quantum threats, providing a safeguard during the critical transition period.\\
In contrast, the second strategy involves using multiple post-quantum algorithms in tandem, drawing on the diversity of different quantum-resistant techniques to bolster security  \cite{ricci2024hybrid}. By combining different approaches, such as lattice-based, hash-based, and code-based schemes, this method mitigates the risk of vulnerabilities inherent in any single algorithm, though it introduces increased computational overhead. Despite the higher resource requirements, this multi-algorithm approach offers stronger security guarantees by diversifying the underlying assumptions and creating more robust defenses against quantum attacks.

\subsection{Practical Considerations}
Practical considerations in implementing PQC involve addressing both the challenges of key and signature size management and optimizing system performance. One of the primary challenges is handling the significantly larger key and signature sizes typical of post-quantum algorithms \cite{fathalla2024beyond}. These increased sizes necessitate protocol modifications to accommodate larger data structures, as well as considerations for storage and transmission, which may strain existing infrastructure. \\ 
Additionally, performance optimization becomes a crucial factor, requiring hardware acceleration opportunities, such as specialized cryptographic processors, and software optimization techniques to enhance processing efficiency \cite{malina2019towards,sood2024cryptography}. Memory management strategies are also essential to ensure that the increased computational demands do not hinder system performance \cite{caruso2024post}. In parallel, security levels defined by NIST provide a framework for evaluating the strength of PQC schemes. These levels, ranging from Level 1 (equivalent to AES-128\footnote{Advanced Encryption Standard
}) to Level 5 (equivalent to AES-256), offer guidelines for selecting appropriate algorithms based on the required security guarantees, ensuring that organizations can align their cryptographic implementations with the necessary protection against evolving quantum threats.

\subsection{Migration Strategies}
Migration strategies for adopting PQC emphasize careful planning and adherence to established standards to ensure a seamless transition. A common approach is gradual migration, starting with non-critical systems to minimize risk \cite{newhouse1800migration}. During this phase, hybrid cryptographic schemes that combine classical and post-quantum algorithms can be used to ensure continuity of security while the transition progresses. Continuous monitoring for implementation issues is essential to identify and address potential challenges early in the process.\\
Testing and validation play a pivotal role in ensuring the success of the migration, including performance testing in real environments to assess the efficiency of new cryptographic solutions, interoperability testing\footnote{Interoperability tests assess performance when standardized biometric data records are used across systems from different vendors.} to ensure compatibility with existing systems, and rigorous security validation procedures to confirm the resilience of the new system against emerging threats \cite{nather2024migrating}. Adherence to standards is critical, particularly following the NIST PQC standardization process, which ensures the implementation of approved algorithms and maintains comprehensive compliance documentation \cite{ott2019identifying}. In addition to NIST standards, staying updated with industry developments, such as those from the TLS working group\footnote{The Transport Layer Security (TLS) working group was formed in 1996 to develop a standardized security protocol for the transport layer.}, X.509 certificate\footnote{An X.509 certificate links an identity to a public key through a digital signature.} updates, and protocol modifications, helps organizations align with best practices and ensure long-term compatibility and security in a post-quantum world \cite{kampanakis2018viability}.

\section{QUASAR Framework and pipeline for transforming and  preparing your organization for post quantum era}  
%
The first step in the QUASAR framework is the assessment and discovery phase. This phase is critical for establishing a comprehensive understanding of an organization’s current cryptographic landscape and identifying areas that are vulnerable to quantum-era threats. A successful quantum-resilient transition begins with complete visibility into where and how cryptographic mechanisms are used across the organization.

\subsubsection{Cryptographic Inventory}
A thorough cryptographic inventory forms the foundation for any post-quantum migration effort. This process involves systematically identifying and documenting all cryptographic assets, dependencies, and implementations within the organization. The key activities in this sub-phase include: Mapping Cryptographic Assets and Dependencies, Documenting Algorithms, Key Sizes, and Protocols, Identifying Certificate Authorities and PKI Infrastructure, Cataloging Encrypted Data Storage, and Surveying Network Security Protocols.

\paragraph{Mapping Cryptographic Assets and Dependencies}
Begin by identifying all systems, applications, and components—both internal and third-party—that utilize cryptographic functions. This includes software libraries, Application Programming Interfaces (APIs), Hardware Security Modules (HSMs), and embedded systems. Dependencies should be analyzed recursively to uncover indirect cryptographic usage through integrated or inherited components.
\paragraph{Documenting Algorithms, Key Sizes, and Protocols}
For each cryptographic instance, record the specific algorithms such as RSA, ECC, AES, key lengths, and associated protocols such as TLS 1.2/1.3, Internet Protocol Security (IPsec), Secure Shell (SSH). Particular attention should be given to public-key cryptography, which is highly susceptible to quantum attacks \cite{kampanakis2023vision}. Understanding current algorithm usage is essential for determining readiness for post-quantum alternatives.
\paragraph {Identifying Certificate Authorities and PKI Infrastructure}
Catalog the organization’s Public Key Infrastructure, including internal and external Certificate Authorities, root and intermediate certificates, and trust hierarchies. This also includes revocation mechanisms such as Certificate Revocation Lists (CRLs) and Online Certificate Status Protocol (OCSP) responders \cite{berbecaru2023evaluation}. The structure and policies governing certificate issuance and lifecycle management must be clearly documented.
\paragraph{Cataloging Encrypted Data Storage}
Identify all locations where data is encrypted at rest—including databases, file systems, backups, and archives. Record the encryption methods used (such as full-disk encryption, database-level encryption), the cryptographic algorithms involved, and the data classification levels. This information is crucial for prioritizing the migration of high-value or long-retention data that may be susceptible to "harvest now, decrypt later" threats \cite{olutimehin2025future}.
\paragraph{Surveying Network Security Protocols}
Assess all network communication channels that rely on cryptographic protection. This includes transport-layer protocols such as TLS, Virtual Private Network (VPN) implementations (such as IPsec, WireGuard \cite{dowling2018cryptographic}), secure shell protocols (such as SSH), and internal service-to-service encryption. Protocol configurations should be examined for algorithm agility and compatibility with quantum-safe alternatives \cite{baseri2024navigating}.\\
By conducting a detailed and structured cryptographic inventory, organizations can establish a baseline of their cryptographic posture. This inventory not only identifies immediate vulnerabilities but also informs a risk-based prioritization strategy for transitioning to quantum-resistant cryptographic solutions. The outputs of this phase serve as essential inputs for the next stage: risk analysis and transition planning.

\subsubsection{Risk Assessment}
Following the cryptographic inventory, the Risk Assessment phase provides a strategic foundation for prioritizing the migration to quantum-resistant cryptographic solutions. This step focuses on evaluating the relative risk posed to various systems, processes, and data types by emerging quantum threats. By integrating technical, operational, and regulatory factors, organizations can develop a risk-informed transition roadmap.
This section covers the following key components of risk assessment: Evaluating Data Sensitivity and Protection Requirements, Assessing the Quantum Threat Timeline for Different Systems, Identifying Critical Business Processes Dependent on Cryptography, Documenting Compliance and Regulatory Requirements, and Analyzing Supply Chain Cryptographic Dependencies.
\paragraph{Evaluating Data Sensitivity and Protection Requirements}
classify all data assets according to their sensitivity, confidentiality, and regulatory status. High-value assets such as Personally Identifiable Information (PII), financial records, intellectual property, and national security data require enhanced protection and should be prioritized for early quantum-resilient upgrades. The required protection duration should also be considered—for instance, data that must remain confidential for a decade or more is particularly vulnerable to long-term quantum threats \cite{malina2021post}.
\paragraph{Assessing the Quantum Threat Timeline for Different Systems}
Not all systems are equally exposed to quantum computing risks. Evaluate the lifespan, exposure, and cryptographic reliance of each system to estimate when a quantum adversary could realistically compromise it. Legacy systems, public-facing services, and assets stored in encrypted archives are especially vulnerable to "harvest now, decrypt later" attacks, where encrypted data is collected today with the intent of decryption once quantum capabilities mature \cite{nather2024migrating}.
\paragraph{Identifying Critical Business Processes Dependent on Cryptography}
Cryptographic functions underpin a wide range of business operations, from secure communication and digital signatures to identity management and data integrity. Identify and map business-critical processes that rely on cryptography. Examples include transaction processing, authentication services, software updates, and supply chain integrity. Disruption to these processes can result in significant operational and reputational damage, making them top priorities in the post-quantum transition strategy.
\paragraph{Documenting Compliance and Regulatory Requirements}
Many sectors are governed by strict compliance mandates related to cryptographic strength, data retention, and breach notification. Review applicable regulations such as General Data Protection Regulation (GDPR), Health Insurance Portability and Accountability Act (HIPAA), Payment Card Data Security Standards (PCI DSS), and sector-specific frameworks \cite{sargiotis2024legal}. Determine which systems are bound by regulatory obligations that may require proactive adoption of quantum-safe standards, once formalized by bodies such as NIST or European Telecommunications Standards Institute (ETSI) \cite{lonc2023feasibility}.
\paragraph{Analyzing Supply Chain Cryptographic Dependencies}
Modern organizations often rely on third-party services, vendors, and partners that implement their own cryptographic solutions. Perform a dependency analysis to identify how and where external providers use cryptography, and evaluate their readiness for the quantum era. Supply chain vulnerabilities can introduce hidden risks, as unprepared vendors may become weak links in otherwise secure architectures.\\

\subsubsection{Impact Analysis}
The Impact Analysis phase translates quantum-related risks into tangible organizational consequences. By quantifying the potential fallout of quantum attacks and forecasting the resource and operational demands of a cryptographic migration, this phase enables informed decision-making and strategic planning \cite{ali2021pragmatic}. It also ensures alignment between technical change and business continuity.\\
The following analysis is structured around five key areas: Calculating the potential costs of quantum attacks, Estimating resource requirements for transition, Assessing operational impacts of migration, Evaluating business continuity implications, and Identifying affected stakeholders.\\
\paragraph{Calculating the Potential Costs of Quantum Attacks} Estimate the financial, reputational, legal, and operational consequences of a successful quantum-enabled breach. This includes costs associated with data exposure, intellectual property theft, regulatory fines, service downtime, customer loss, and brand damage. For sectors with high-value targets or sensitive data (such as finance, healthcare, defense), the projected costs may justify immediate mitigation efforts and early adoption of quantum-safe solutions.
\paragraph{Estimating Resource Requirements for Transition}
Forecast the human, technological, and financial resources required to plan and implement a quantum-resilient cryptographic infrastructure. This includes software and hardware upgrades, staff training, testing environments, migration tools, and vendor engagement. A realistic resource projection helps scope the transition effort and informs budgeting and timeline decisions.\\
\paragraph{Assessing Operational Impacts of Migration}
Analyze the potential disruptions caused by cryptographic changes to ongoing business operations. This may include service downtime during key transitions, compatibility issues with legacy systems, or performance overhead from new cryptographic algorithms. A phased and well-tested deployment plan is critical to minimizing adverse operational impacts. 
\paragraph{Evaluating Business Continuity Implications} Review the organization’s business continuity and disaster recovery strategies in the context of quantum preparedness. Ensure that contingency plans address potential failures in cryptographic functions, such as authentication outages or certificate validation errors, that could arise during or after the transition. Consider testing business continuity plans under post-quantum conditions to ensure resilience. 
\paragraph {Identifying Affected Stakeholders}
Map the internal and external stakeholders impacted by the cryptographic transition. Internally, this may include IT, security, compliance, legal, operations, and executive leadership. Externally, it encompasses customers, partners, regulators, and vendors. Clear communication and coordinated planning across stakeholders are essential for ensuring support, compliance, and a smooth transition.

\subsection{Strategy Development}
Once risks and impacts have been assessed, the next step is to define a structured and coordinated approach to transformation. Strategy Development involves establishing the organizational, procedural, and decision-making foundations required to guide the transition toward quantum-resilient cryptography \cite{aydeger2024towards,joseph2022transitioning}. A strong governance structure is essential to ensure accountability, alignment, and sustained momentum throughout the multi-year transformation process \cite{pandeya2021strategy}.
\subsubsection{Governance Structure}
A well-defined governance model ensures that the post-quantum transition is managed consistently, strategically, and in alignment with organizational priorities \cite{alghamdi2025post}. This structure should provide leadership, oversight, and cross-functional coordination from the outset.
This section addresses Establishing a post-quantum steering committee, Defining roles and responsibilities, Creating reporting and oversight mechanisms, Developing decision-making frameworks, and Setting up communication channels.
\paragraph {Establishing a Post-Quantum Steering Committee}
Create a dedicated steering committee to lead and oversee the PQC transformation. This multidisciplinary group should include representatives from information security, IT, compliance, legal, risk management, operations, and executive leadership. The committee will be responsible for setting the strategic direction, aligning resources, and monitoring progress across workstreams.
\paragraph {Defining Roles and Responsibilities}
Clearly define roles and responsibilities across the organization to avoid gaps or redundancies. Assign owners for specific initiatives such as cryptographic migration, vendor engagement, regulatory alignment, and training. Responsibility matrices (such as RACI charts\footnote{The acronym RACI stands for Responsible, Accountable, Consulted, and Informed. is a framework used to define the roles and level of involvement of different individuals or groups in completing tasks or deliverables within a project or business process.}) can support transparency and accountability. 
\paragraph {Creating Reporting and Oversight Mechanisms} Establish a system for regular reporting, including status updates, Key Performance Indicators (KPIs), risk tracking, and milestone reviews. These mechanisms ensure that progress is measurable and deviations are addressed early. Oversight bodies such as audit committees or cybersecurity risk boards may also be involved in periodic reviews.
\paragraph {Developing Decision-Making Frameworks}
Define the decision-making processes for technology adoption, investment approval, exception handling, and vendor selection. Decisions should be guided by predefined criteria based on risk impact, compliance requirements, and business priorities. A formal escalation path should be included for high-impact or cross-functional decisions.
\paragraph{Setting Up Communication Channels}
Effective communication is critical for organizational alignment and stakeholder engagement. Establish internal communication channels—such as regular briefings, dashboards, or newsletters—to keep teams informed. External communication strategies may also be required to coordinate with regulators, partners, and customers, particularly when public-facing services are affected.

\subsubsection{Policy Framework}
A robust Policy Framework is essential to formalize organizational intent and ensure consistency in the adoption of PQC principles \cite{dekker2022regulating}. Policies act as the binding force between strategy and execution, aligning technical initiatives with security objectives, regulatory obligations, and operational practices. This framework must evolve to support both the transition period and long-term maintenance of quantum-resilient systems.\\
This section covers Defining post-quantum security policies, Establishing crypto-agility requirements, Creating procurement guidelines, Developing compliance frameworks, and Setting security standards.
\paragraph{Defining Post-Quantum Security Policies}
Develop formal security policies that articulate the organization's commitment to post-quantum readiness. These should define the scope of cryptographic modernization efforts, mandate risk-based prioritization, and specify long-term goals such as eliminating quantum-vulnerable algorithms. Policies should also reflect evolving standards from authoritative bodies like NIST, ETSI, and national cybersecurity agencies.
\paragraph{Establishing Crypto-Agility Requirements}
To future-proof cryptographic systems, define requirements for crypto-agility—the ability to rapidly switch cryptographic algorithms and protocols without significant architectural changes. Policy mandates should require the design and procurement of systems that support algorithm agility, version control, and centralized key management to accommodate future cryptographic shifts, including hybrid or layered approaches.
\paragraph{Creating Procurement Guidelines}
Update procurement policies to include post-quantum security considerations. Vendors and third-party providers should be evaluated on their cryptographic transparency, roadmap for PQC support, and adherence to recognized standards. Procurement guidelines should require suppliers to disclose cryptographic algorithms used, support quantum-safe options, and provide a transition timeline aligned with the organization’s goals.
\paragraph{Developing Compliance Frameworks}
Integrate post-quantum readiness into existing compliance frameworks to ensure regulatory alignment and audit readiness. This includes mapping quantum transition requirements to existing standards (such as ISO 27001, PCI-DSS, GDPR) and developing metrics to track conformance \cite{adeyinka2025cybersecurity}. Where regulations are still evolving, prepare to adopt future requirements through adaptive policy mechanisms and continuous monitoring.
\paragraph{Setting Security Standards}
Establish internal cryptographic standards that define approved algorithms, key sizes, and protocol configurations. As NIST finalizes PQC algorithm recommendations, these standards should be updated accordingly. Include guidance on hybrid cryptography, key management, certificate lifecycle, and acceptable use of legacy algorithms during the transition period. Standards should also be tailored for different environments, such as cloud, edge, mobile, and IoT systems.

\subsection{Resource Planning}
A successful transition to PQC requires careful and proactive Resource Planning. This phase ensures that the organization is equipped—financially, technically, and operationally—to carry out the necessary changes over time \cite{nather2024migrating}. Planning for adequate resources helps mitigate transition risks, avoid costly delays, and maintain momentum throughout the migration process. \\
This section addresses Budget allocation, Staffing requirements, Training needs assessment, Technology requirements, and External expertise needs.
\paragraph{Budget Allocation} Allocate financial resources to cover all phases of the quantum-readiness initiative, including assessment, technology upgrades, testing, training, and external advisory support. Budgeting should account for both capital expenditures such as hardware replacements, tooling and operational costs such as maintenance, ongoing assessments, and licensing. Multi-year budgeting strategies may be necessary given the extended timeline of the post-quantum transition. 
\paragraph{Staffing Requirements} Determine the human resource needs for the project, identifying required roles such as cryptographic engineers, cybersecurity architects, compliance specialists, and project managers. In many organizations, PQC implementation will span multiple departments, requiring dedicated coordination to avoid bottlenecks. Resourcing plans should ensure sufficient coverage for both implementation and operational continuity.
\paragraph{Training Needs Assessment} Assess the current level of cryptographic literacy across key teams and identify gaps related to post-quantum topics. Develop a training program to upskill staff in areas such as quantum threats, post-quantum algorithms, crypto-agility, and secure system design. Training may be tiered based on job functions, with in-depth technical workshops for developers and high-level awareness sessions for executives and stakeholders.
\paragraph{Technology Requirements} Evaluate the tools, platforms, and infrastructure needed to support the quantum-resilient transformation. This includes cryptographic libraries with PQC support, test environments, algorithm benchmarking tools, certificate management systems, and crypto-agile frameworks. Technology requirements should align with architectural goals, scalability needs, and compliance obligations.
\paragraph{External Expertise Needs} Identify areas where third-party expertise is required, such as cryptographic consulting, standards alignment, or vendor evaluation. External support may also be necessary for independent assessments, algorithm validation, and integration of specialized tools. Partnerships with academia, national institutes, or industry consortiums such as NIST PQC working groups can also provide valuable insights and early access to emerging best practices.

\subsection{Technical Foundation}
The Technical Foundation phase focuses on establishing the architectural and infrastructural underpinnings necessary to support a secure and efficient transition to PQC \cite{adapa2025architecting}. This includes building systems that are not only secure against quantum threats but also adaptable to the evolving cryptographic landscape. Careful planning and design at this stage help prevent costly rework and ensure long-term maintainability.

\subsubsection{Architecture Development}
Developing a resilient and forward-looking technical architecture is a cornerstone of quantum-readiness. This sub-phase emphasizes modularity, agility, and interoperability to accommodate both classical and PQC algorithms during the transition period  \cite{fernandez2019pre}.\\
Key focus areas in this phase include Designing crypto-agile infrastructure, Planning hybrid classical/post-quantum systems, Developing migration architecture, Creating testing environments, and Designing monitoring systems.
\paragraph{Designing Crypto-Agile Infrastructure}
Design infrastructure that supports crypto-agility, enabling seamless updates or replacements of cryptographic algorithms with minimal disruption. This includes abstracting cryptographic services through interfaces and APIs, centralizing key and certificate management, and avoiding hardcoded cryptographic primitives. Crypto-agile design ensures the organization can respond rapidly as new standards emerge or as vulnerabilities are discovered. 
\paragraph{Planning Hybrid Classical/Post-Quantum Systems}
During the transition period, hybrid cryptographic schemes—combining classical and post-quantum algorithms will play a crucial role in maintaining backward compatibility while strengthening security. Develop architectural patterns that support hybrid deployments, ensuring that systems can validate and interoperate with both legacy and quantum-resistant components. These designs must consider key negotiation, signature validation, and protocol adaptation.
\paragraph{Developing Migration Architecture}
Construct a migration architecture that defines how legacy systems will be upgraded or phased out. This includes designing pathways for gradual replacement of vulnerable components, managing data re-encryption, and ensuring service continuity. The migration architecture should also account for critical dependencies, system interoperability, and prioritization based on risk and business impact.
\paragraph{Creating Testing Environments} 
Establish dedicated environments for testing post-quantum algorithms, cryptographic libraries, and integration with existing systems. These environments should simulate real-world workloads and include test harnesses, benchmarking tools, and security analysis frameworks. Early testing enables validation of performance, interoperability, and correctness, reducing deployment risks in production environments.
\paragraph{Designing Monitoring Systems}
Develop monitoring capabilities to track cryptographic usage, identify deprecated algorithms, and detect anomalies in cryptographic operations. Instrument systems with telemetry to gain visibility into algorithm adoption and system behavior. Monitoring is essential for managing the ongoing transition, ensuring compliance with internal standards, and identifying areas requiring remediation.

\subsubsection{Standards Selection}
As the foundation for secure and interoperable cryptographic systems, Standards Selection plays a central role in ensuring that post-quantum implementations are technically sound, future-proof, and aligned with global best practices \cite{joseph2022transitioning,shamo2024bridging}. This phase focuses on selecting algorithms and defining implementation criteria based on emerging standards, performance considerations, and security objectives.\\
Topics explored in this section include Monitoring NIST PQC standards, Selecting appropriate algorithms, Defining implementation standards, Establishing performance benchmarks, and Creating testing criteria.
\paragraph{Monitoring NIST PQC Standards}
Closely track the progress of the NIST PQC Standardization Project, which is actively finalizing a set of algorithms for quantum-resistant encryption and digital signatures. Organizations should stay informed about algorithm maturity, standardization milestones, and any cryptanalysis findings. Participation in or alignment with NIST’s process ensures compatibility with widely accepted industry and government expectations.\\
\paragraph{Selecting Appropriate Algorithms} 
Based on use cases and threat models, select quantum-resistant algorithms that meet the organization’s security, performance, and integration requirements. Consider factors such as algorithm category such as lattice-based, and code-based, cryptographic function (key exchange, digital signatures), key sizes, and computational overhead. Initial selections may include NIST’s Round 3 finalists such as CRYSTALS-Kyber for key encapsulation and CRYSTALS-Dilithium for digital signatures, with ongoing evaluation of emerging candidates. \\
\paragraph{Defining Implementation Standards}
Establish internal implementation standards to guide secure and consistent adoption of PQC algorithms. These standards should address secure parameter selection, key and certificate handling, interface specifications, and safe fallback mechanisms. They should also define coding practices, input validation, and protections against side-channel attacks. Consistent implementation reduces fragmentation and supports long-term maintainability. \\
\paragraph{Establishing Performance Benchmarks}
Define performance benchmarks to evaluate the efficiency of post-quantum algorithms in various environments, including endpoints, servers, embedded systems, and cloud infrastructures. Benchmarks should measure key generation, encryption/decryption, signing/verification, memory usage, and communication overhead. This helps inform algorithm selection and identifies potential bottlenecks early in the deployment process. \\
\paragraph{Creating Testing Criteria}
Develop rigorous testing criteria to validate the correctness, security, and interoperability of post-quantum implementations. Testing should include unit tests, conformance tests, regression tests, and fuzzing for edge-case behavior. Where applicable, leverage open-source test suites and reference implementations provided by standards bodies. Testing criteria must be integrated into the development lifecycle and continuously updated as standards evolve.

\subsubsection{Pilot Environment}
The Pilot Environment phase serves as a critical bridge between design and deployment, allowing organizations to validate their PQC strategies in controlled, low-risk settings. By building and testing early-stage implementations, organizations can identify technical challenges, evaluate performance impacts, and refine integration strategies before scaling to production. \\
This phase involves Building test infrastructure, Implementing selected algorithms, Developing proof of concepts, Testing performance impacts, and Validating integration approaches.
\paragraph{Building Test Infrastructure}
Establish a dedicated and isolated infrastructure for conducting pilot deployments of PQC systems. This environment should closely mirror production conditions, including representative workloads, network topologies, authentication systems, and data types. Virtualized or containerized testbeds can provide flexibility and scalability for evaluating multiple configurations and use cases.\\
\paragraph{Implementing Selected Algorithms}
Integrate the post-quantum algorithms selected during the standards selection phase into test systems and applications. Use validated cryptographic libraries that support NIST finalist algorithms (such as CRYSTALS-Kyber, CRYSTALS-Dilithium) and ensure that implementations adhere to internal security standards and parameter configurations. Emphasis should be placed on real-world integration with protocols such as TLS, SSH, VPNs, and PKI infrastructures.\\
\paragraph{Developing Proof of Concepts (PoC)}
Create PoC implementations to demonstrate feasibility and explore integration scenarios. PoCs can span various domains—such as secure email, encrypted messaging, authentication workflows, or digital signature services—and help build institutional knowledge while surfacing practical considerations like library compatibility, protocol negotiation, and user experience. \\
\paragraph{Testing Performance Impacts}
Measure and analyze the performance implications of PQC operations on system components. KPIs include latency, throughput, memory usage, and Central Processing Unit (CPU) consumption. Comparative analysis against classical algorithms can inform optimization strategies and help determine where hybrid models or hardware acceleration may be needed to maintain acceptable service levels. \\
\paragraph{Validating Integration Approaches}
Evaluate the effectiveness of integration strategies with existing infrastructure and services. This includes assessing protocol compatibility, Application Programming Interface (API) behavior, key management flows, and certificate handling. Particular attention should be given to backward compatibility and the coexistence of classical and post-quantum algorithms in hybrid environments. Successful validation ensures a smoother migration path to production environments.

\subsection{Implementation Pipeline}
The Implementation Pipeline translates strategic plans and technical designs into actionable initiatives that lead to the deployment of quantum-resilient cryptographic systems \cite{kappler2022post,xie2020special}. This phase is structured to ensure that migration activities are executed methodically, with minimal disruption to business operations. 

\subsubsection{Preparation Stage}
The Preparation Stage ensures that all organizational, technical, and procedural elements are in place before executing cryptographic transitions. By aligning tools, teams, environments, and communication channels, organizations reduce risks and improve coordination during implementation.\\
Key activities in this stage include Tool selection and setup, Environment preparation, Team training, Documentation setup, and Communication planning
\paragraph{Tool Selection and Setup}
Identify and configure the software tools, libraries, and platforms required for the deployment of post-quantum algorithms. This includes selecting cryptographic libraries with PQC support, integration frameworks, testing tools, certificate management systems, and automated deployment pipelines. Tools should be validated for compatibility with selected algorithms and adherence to internal security policies. \\
\paragraph{Environment Preparation}
Prepare development, staging, and production environments for the deployment of post-quantum components. Ensure that infrastructure is capable of supporting increased computational demands and hybrid algorithm configurations. This may include upgrading hardware, provisioning virtual environments, or establishing secure testing sandboxes for pre-deployment validation.\\
\paragraph{Team Training}
Equip technical and operational teams with the knowledge needed to execute the migration. This includes targeted training on post-quantum algorithms, crypto-agile architectures, deployment procedures, monitoring tools, and incident response protocols. Training ensures teams can identify, troubleshoot, and resolve issues that may arise during implementation. \\
\paragraph{Documentation Setup}
Develop and organize detailed documentation to support the transition process. This includes architecture blueprints, deployment runbooks, change management procedures\footnote{Change management focuses on guiding organizational transitions by implementing strategies that prepare and support individuals, teams, and leaders through the change process.}, configuration guides, and rollback plans. Clear documentation ensures knowledge transfer, enables repeatability, and serves as a reference for audits and future updates. \\
\paragraph{Communication Planning} 
Establish internal and external communication plans to support transparency and coordination throughout implementation. Define communication channels, update schedules, stakeholder responsibilities, and escalation paths. Communication planning also ensures that executives, end-users, partners, and regulatory bodies are informed of planned changes, potential impacts, and support procedures.

\subsubsection{Development Stage}
The Development Stage is where the technical transition to PQC takes form \cite{gharavi2024post,algazy2024development}. This stage focuses on the hands-on implementation of selected cryptographic algorithms and the modification of system components to ensure compatibility, performance, and security. It is a highly iterative process involving design adaptation, testing, and validation.\\
Development activities include Algorithm implementation, Code Development, Integration testing, Performance optimization, Security validation, Documentation, System updates, Protocol modifications, API updates, Database changes, Network configurations, and Security controls.
\paragraph{Algorithm Implementation}
This sub-stage focuses on embedding quantum-resistant algorithms into existing systems and applications. The objective is to ensure that cryptographic operations are executed securely and efficiently, while maintaining interoperability with other systems.\\
\paragraph{Code Development} Develop or integrate PQC algorithms into application codebases using vetted libraries and Software Development Kits (SDKs). Ensure secure coding practices are followed, and abstract cryptographic operations to facilitate future updates or algorithm replacements.\\
\paragraph{Integration Testing}
Conduct comprehensive testing to validate that new cryptographic components function correctly within system workflows. This includes key exchange, authentication, and data encryption/decryption processes. Test for compatibility with existing protocols and legacy components where hybrid approaches are used.\\
\paragraph{Performance Optimization}
Analyze and optimize the performance of post-quantum implementations to mitigate increased computational or bandwidth overhead. Apply optimizations at the algorithm level, as well as in surrounding application logic, caching mechanisms, and data flow architectures.\\
\paragraph{Security Validation}
Perform security testing to identify and remediate vulnerabilities such as side-channel leaks, input validation flaws, and incorrect parameter usage. Use static analysis, dynamic testing, and third-party reviews to ensure robustness against classical and emerging quantum threats.\\
\paragraph{Documentation}
Document implementation decisions, code-level integrations, algorithm parameters, and performance results. Maintain clear developer guides and internal references to support future audits, refactoring, or algorithm migrations.\\
\paragraph{ System Updates}
Adapting the broader system environment to support PQC is essential for successful deployment. This involves modifying protocols, interfaces, and system configurations to accommodate new cryptographic operations.\\
\paragraph{Protocol Modifications}
Update communication protocols (such as TLS, SSH, IPsec) to support quantum-safe key exchange and digital signature schemes. Implement protocol negotiation mechanisms to enable coexistence with legacy systems during the migration period.\\
\paragraph{API Updates}
Modify internal and external APIs to support updated cryptographic functions, new key formats, and certificate handling procedures. Ensure backward compatibility is preserved where necessary to minimize disruption to clients and services.\\
\paragraph{Database Changes}
Adjust database schemas and storage logic to support longer keys, expanded certificate fields, or additional metadata required by post-quantum operations. Ensure encryption and hashing procedures in data at rest are updated accordingly.\\
\paragraph{Network Configurations}
Update network infrastructure—such as firewalls, load balancers, and VPN configurations—to support revised encryption standards, increased payload sizes, and new authentication mechanisms. Validate end-to-end encryption paths.\\
\paragraph{Security Controls}
Review and enhance access control, audit logging, and intrusion detection systems to accommodate new cryptographic primitives and potential changes in threat models. Ensure post-quantum deployments are fully integrated into the organization’s security monitoring framework.

\subsubsection{Testing Stage}
The Testing Stage is critical to ensure that all PQC implementations function as expected and meet security, performance, and compliance standards \cite{alagic2019status,joseph2022transitioning}. This phase involves systematic testing and validation to confirm that the system meets both functional and non-functional requirements before deployment in production environments.\\
Testing activities include Functional testing, Unit tests, Integration tests, System tests, Performance tests, Security tests, Validation, Compliance checking, Standards verification, Performance validation, Security assessment, and Documentation review.
\paragraph{Functional Testing}
Functional testing ensures that the post-quantum algorithms and system modifications operate as intended and integrate seamlessly with other system components. This testing focuses on correctness and operational readiness.\\
\paragraph{Unit Tests}
Develop unit tests for individual cryptographic functions to verify that core components—such as key generation, encryption/decryption, and digital signatures—work correctly in isolation. Unit tests should cover edge cases and boundary conditions, ensuring robustness under various input scenarios.\\
\paragraph{Integration Tests}
Conduct integration tests to ensure that cryptographic components interact properly with other systems and services. This includes testing cryptographic protocols such as TLS and SSH with the new quantum-safe algorithms and verifying that the system integrates smoothly with external services, databases, and networks.\\
\paragraph{System Tests}
Perform end-to-end testing on complete system workflows that involve cryptographic operations. This verifies that the overall system functions as intended when cryptographic elements are incorporated, ensuring proper behavior in typical usage scenarios and across various application layers.\\
\paragraph{Performance Tests} 
Execute performance testing to measure the impact of post-quantum algorithms on system response times, throughput, and resource utilization. Key metrics include latency, memory consumption, CPU usage, and bandwidth requirements. This helps identify any bottlenecks or areas requiring optimization. \\
\paragraph{Security Tests}
Conduct security testing to identify vulnerabilities that may arise from the introduction of new cryptographic components. This includes testing for potential side-channel attacks, incorrect parameter handling, and weaknesses in key management or protocol implementations. Use tools like fuzzing, penetration testing, and static code analysis to uncover hidden issues.
\paragraph{Validation}
Validation ensures that the system meets external and internal standards, regulations, and performance expectations. It confirms that the implemented cryptographic solutions align with security objectives, industry best practices, and compliance requirements.\\
\paragraph{Compliance Checking} Verify that the PQC implementation complies with relevant legal, regulatory, and industry standards (such as GDPR, PCI DSS, HIPAA). This includes reviewing data protection measures, key management procedures, and reporting protocols to ensure that all compliance obligations are met.\\
\paragraph{Standards Verification}
Ensure that the system adheres to recognized standards and guidelines, such as those provided by NIST, ETSI, and ISO. This includes verifying that the selected algorithms are properly implemented, and that cryptographic modules are certified or validated according to applicable standards.\\
\paragraph{Performance Validation}
Validate that performance benchmarks meet the required service levels under real-world operating conditions. Compare actual performance results with pre-defined performance objectives, ensuring that the system can handle expected workloads without introducing excessive latency or resource consumption. \\
\paragraph{Security Assessment}
Perform a comprehensive security assessment to evaluate the overall resilience of the system. This includes vulnerability scans, threat modeling, and risk assessments focused on the newly integrated cryptographic components. Independent third-party audits may also be considered to validate the robustness of the security posture. \\
\paragraph{Documentation Review}
Review all documentation related to the cryptographic implementation, including design specifications, code comments, test plans, and deployment instructions. Ensure that the documentation is complete, accurate, and up to date, providing clear guidance for future updates, audits, and troubleshooting.

\subsubsection{Deployment Stage}
The Deployment Stage marks the transition of PQC components from controlled testing environments to operational systems \cite{geremew2024preparing}. It is a critical phase that requires careful coordination, thorough planning, and robust fallback mechanisms to ensure a stable and secure deployment. \\
Key components of this stage include Rollout planning, Phasing strategy, Backup procedures, Rollback plans, Monitoring setup, Support preparation, Implementation, System deployment, User training, Performance monitoring, Issue tracking, and Documentation updates.
\paragraph{Rollout Planning}
Effective rollout planning minimizes risks and ensures business continuity during the deployment of post-quantum solutions. A strategic, phased approach is essential for managing complexity and adapting to unforeseen issues.\\
\paragraph{Phasing Strategy} Define a phased rollout plan that prioritizes high-risk or high-value systems and progressively expands to lower-priority assets. Rollout may be conducted by business unit, geography, or technology stack. This staged approach enables early feedback, reduces the blast radius of potential failures, and supports continuous refinement. \\
\paragraph{Backup Procedures}
Ensure reliable and current backups are in place before any cryptographic changes are applied. Backup plans should cover data, configuration states, certificates, keys, and system images. Validate backup integrity and Recovery Time Objectives (RTO)\footnote{The Recovery Time Objective (RTO) defines the maximum allowable duration that an application, system, network, or service can remain unavailable following an unexpected disruption such as a disaster or system failure.} through pre-deployment testing. \\
\paragraph{Rollback Plans}
Develop and document rollback procedures to restore systems to a known good state in case of deployment failure. This includes clear instructions for reverting cryptographic changes, recovering data, and reinitializing services without compromising security or functionality.\\
\paragraph{Monitoring Setup}
Deploy monitoring tools to observe system behavior, cryptographic operations, and network traffic post-deployment. Monitoring should provide real-time visibility into performance metrics, error rates, and potential anomalies introduced by the new cryptographic components.\\
\paragraph{Support Preparation}
Prepare helpdesk and support teams for deployment-related inquiries, incidents, and escalations. Provide training, runbooks, and escalation paths to handle issues quickly and efficiently. Establish feedback loops for reporting issues back to development and deployment teams.
\paragraph{Implementation}
The implementation sub-phase involves the actual application of cryptographic changes to production environments. It must be executed with precision, validated in real-time, and supported by comprehensive user and system readiness measures.\\
\paragraph{System Deployment} 
Execute the deployment of PQC libraries, services, certificates, and configurations across targeted systems. This includes ensuring all dependencies are updated and that systems are correctly configured for hybrid or quantum-safe operation modes. 
\paragraph{User Training}
Provide targeted training for end-users, administrators, and stakeholders who interact with systems affected by the cryptographic changes. Training may include updates to authentication flows, secure communication practices, and awareness of new tools or interfaces. \\
\paragraph{Performance Monitoring}
Continuously monitor system performance to detect any degradation caused by new cryptographic operations. Compare observed metrics to pre-deployment baselines and adjust configurations or optimizations as needed to maintain service levels. \\
\paragraph{Issue Tracking}
Implement a centralized issue tracking process to log and manage any errors, regressions, or user-reported problems during the deployment. Ensure issues are triaged, prioritized, and resolved promptly with input from cross-functional teams. \\
\paragraph{Documentation Updates}
Update all operational, technical, and user documentation to reflect the newly deployed cryptographic systems. This includes configuration guides, Standard Operating Procedures (SOPs), architecture diagrams, and policy documents. Keeping documentation accurate ensures alignment across teams and simplifies future audits or upgrades.

\subsection{Operational Integration}
The Operational integration phase ensures that the deployed PQC solutions are continuously monitored, managed, and maintained throughout their lifecycle \cite{nookala2024post,hasan2024framework}. This phase is critical to identify potential issues, ensure optimal performance, and maintain compliance with security and regulatory standards.

\subsubsection{Monitoring Framework}
A robust Monitoring Framework is essential for maintaining the health and security of PQC systems. Continuous monitoring provides insights into system performance, detects potential vulnerabilities, and ensures compliance with established policies. This stage involves defining key monitoring activities across various dimensions of the deployed systems.\\
This phase comprises several key focus areas, namely Performance metrics, Security monitoring, Compliance tracking, Issue detection, and Usage analysis.
\paragraph{Performance Metrics} 
Establish and track performance metrics to ensure that PQC algorithms and systems perform within acceptable parameters. KPIs should include response times for encryption and    decryption, key generation time, network throughput, memory usage, and CPU consumption. Benchmarking these metrics ensures that the system remains efficient and meets operational expectations. Regular performance reviews can also highlight areas for optimization or hardware scaling.\\
\paragraph{Security Monitoring} 
Implement continuous security monitoring to detect and respond to potential vulnerabilities or threats targeting the cryptographic systems. This includes monitoring for unusual patterns in encryption operations, unauthorized access attempts, side-channel attacks, and anomalies in key management processes. Security monitoring should be integrated with existing Security Operations Centers (SOCs) and leverage automated detection tools, such as Intrusion Detection Systems (IDS) and anomaly detection models, to quickly identify potential risks. \\
\paragraph{Compliance Tracking} Regularly track and validate compliance with regulatory requirements and industry standards such as GDPR, PCI DSS, and HIPAA as they pertain to cryptographic operations. Automated compliance checks should be implemented to ensure that cryptographic components and key management practices align with legal and organizational requirements. The monitoring system should generate alerts for any deviations from established compliance policies, ensuring timely corrective actions. \\
\paragraph{Issue Detection}
Set up automated alerting systems for the detection of operational issues, such as failed encryption attempts, expired certificates, or misconfigured cryptographic settings. These alerts should be integrated with issue tracking systems to provide detailed diagnostic information and facilitate quick resolution. Proactive issue detection helps prevent system downtime or security incidents by addressing concerns before they escalate. \\
\paragraph{Usage Analysis}
Conduct regular usage analysis to assess how cryptographic systems are being utilized across the organization. This includes tracking the frequency of cryptographic operations, identifying heavily used encryption functions, and understanding the operational impact of these functions on business processes. Usage data can also inform future capacity planning, enabling the organization to scale infrastructure in anticipation of increased demand or evolving cryptographic standards.

\subsubsection{Maintenance Procedures}
Effective Maintenance Procedures are essential for ensuring the long-term resilience, performance, and security of PQC systems \cite{aydeger2024towards}. As cryptographic standards continue to evolve and quantum capabilities advance, organizations must adopt proactive maintenance practices to adapt to change while minimizing operational risk. \\
Key areas addressed in this stage include Regular updates, Security patches, Performance optimization, Configuration management, and Documentation maintenance.
\paragraph{Regular Updates} Establish a structured schedule for updating cryptographic libraries, protocols, and related software components. Updates should include new algorithm versions, improved parameter sets, and optimizations from standards bodies such as NIST. Maintenance windows should be defined to allow controlled and minimally disruptive rollouts. Where possible, automated update mechanisms can help reduce administrative overhead while maintaining consistency across systems.\\
\paragraph{Security Patches} Apply security patches promptly to address vulnerabilities in cryptographic implementations, underlying libraries, or system components. Patch management should follow a defined process that includes impact assessment, prioritization, testing, and documented deployment. Given the evolving threat landscape, especially in the quantum context, organizations must remain vigilant and responsive to both zero-day threats and known exploits. \\
\paragraph{Performance Optimization} Continuously monitor and analyze the performance of post-quantum systems to identify opportunities for optimization. This may involve tuning cryptographic parameters, optimizing code paths, improving resource allocation, or upgrading hardware components. Periodic performance audits help ensure that quantum-safe cryptographic operations remain efficient, especially as usage patterns shift or system loads increase. \\ 
\paragraph{Configuration Management} Implement configuration management controls to ensure that cryptographic settings remain consistent, secure, and aligned with policy. This includes managing key lengths, cipher suites, protocol versions, and access permissions. All changes to cryptographic configurations should be tracked through version control systems or configuration management tools to ensure traceability and auditability. \\
\paragraph{Documentation Maintenance} Keep all related documentation current and accessible. This includes architecture diagrams, implementation guides, change logs, configuration files, SOPs, and security policies. As updates and optimizations are introduced, corresponding documentation must be revised to reflect the latest state of the system. Well-maintained documentation supports continuity, compliance, training, and incident response.

\subsubsection{Incident Response}
A robust Incident response capability is essential for effectively managing disruptions or security events involving PQC systems \cite{gbajadesigning,geremew2024preparing}. Given the evolving nature of cryptographic threats, especially in the context of quantum computing, organizations must be prepared to detect, contain, and recover from incidents quickly while minimizing business disruption and reputational impact. \\
The following sections outline the core components of incident response: Detection procedures, Response protocols, Recovery processes, Communication plans, and Lesson learning.
\paragraph{Detection Procedures} Establish advanced monitoring and detection mechanisms tailored to cryptographic anomalies and threats. These should include real-time alerts for unusual key usage patterns, failed cryptographic operations, protocol deviations, unauthorized access to key material, and system integrity violations. Integrating machine learning or behavior-based analytics can enhance early detection of novel or sophisticated attacks, including potential quantum-enabled intrusions. \\ 
\paragraph{Response Protocols} Develop and document standardized response protocols that outline the step-by-step procedures for handling cryptographic incidents. Protocols should include initial triage, severity classification, containment actions, escalation paths, and forensic data collection. Clearly defined roles and responsibilities for security teams, IT personnel, and executive leadership help ensure a coordinated and efficient response. \\
\paragraph{Recovery Processes} Implement recovery processes to restore cryptographic services and secure operations following an incident. This may involve revoking and reissuing certificates, regenerating keys, reverting to known-good configurations, or replacing compromised libraries. Recovery plans must be tested regularly to verify their effectiveness and to ensure minimal downtime during actual incidents. \\
\paragraph{Communication Plans} Create internal and external communication plans to guide information sharing during a cryptographic incident. These plans should specify when and how to notify stakeholders, employees, customers, partners, and regulators. Clear, timely, and accurate communication helps maintain trust and ensures compliance with legal and contractual obligations. \\
\paragraph{Lesson Learning} Conduct post-incident reviews to analyze the root cause, assess the effectiveness of the response, and identify areas for improvement. Lessons learned should be documented and used to update policies, training materials, detection mechanisms, and recovery plans. Continuous learning ensures the organization becomes progressively more resilient with each incident.

\subsection{Continuous Improvement}
Transitioning to PQC is not a one-time effort but an ongoing process that must adapt to emerging threats, evolving standards, and organizational needs \cite{ott2019identifying}. The Continuous improvement phase ensures that the organization remains agile, resilient, and prepared for future developments by embedding feedback loops, reviews, and strategic updates into operational practices.

\subsubsection{Review Process}
A structured review process is critical for maintaining the effectiveness, security, and compliance of PQC systems over time \cite{fathalla2024beyond}. This process involves regularly evaluating all aspects of the post-quantum deployment to identify gaps, optimize performance, and align with new business and regulatory requirements.\\
This review process is structured around five key activities: Regular assessments, Performance reviews, Security audits, Compliance checks, and Stakeholder feedback.
\paragraph{Regular Assessments} Conduct periodic assessments of the post-quantum strategy, architecture, and implementation. These evaluations should address system integrity, algorithm relevance, and the overall alignment of the cryptographic infrastructure with evolving quantum threat landscapes. Scheduled assessments help organizations stay ahead of technological shifts and identify components that may require enhancement or replacement. \\
\paragraph{Performance Reviews} Review system performance metrics regularly to ensure that cryptographic operations continue to meet expected service levels. Analyze trends in processing times, resource utilization, and system responsiveness. These insights support ongoing optimization efforts and guide decisions on infrastructure scaling, algorithm tuning, or hybrid deployment strategies. \\
\paragraph{Security Audits} Perform in-depth security audits to examine the cryptographic environment for vulnerabilities, misconfigurations, or deviations from best practices. These audits should be both internal and third-party, and include assessments of key management, certificate handling, access control, and incident response readiness. Audit findings should be tracked, remediated, and used to inform future policy updates.\\
\paragraph{Compliance Checks} Routinely validate that the organization remains compliant with applicable regulations, industry standards, and internal policies. As legal and regulatory frameworks evolve to address post-quantum risks, compliance checks ensure that cryptographic controls and documentation are up to date and aligned with new requirements. Automated tools and frameworks can assist in maintaining a continuous compliance posture.\\
\paragraph{Stakeholder Feedback} Gather feedback from stakeholders across technical, operational, and business domains to gain insight into how the post-quantum transformation is impacting the organization. Feedback mechanisms should include surveys, interviews, retrospectives, and cross-functional reviews. Incorporating stakeholder input helps refine strategies, address usability concerns, and strengthen organizational buy-in.

\subsubsection{Optimization}
The Optimization phase focuses on refining the PQC systems to ensure they remain efficient, cost-effective, and capable of meeting evolving business and technological demands. This stage involves ongoing efforts to enhance system performance, optimize resource utilization, streamline processes, and reduce operational costs—all while maintaining the security and integrity of cryptographic functions.\\
The following focus areas guide this phase: Performance tuning, Resource optimization, Process improvement, Cost reduction, and Efficiency enhancement.
\paragraph{Performance Tuning} Regular performance tuning is crucial to ensure that PQC systems continue to deliver high throughput and low latency, even as the organization scales. This involves fine-tuning algorithm parameters, optimizing cryptographic protocols, and adjusting system configurations. Performance tests should be conducted periodically to detect and resolve any degradation, ensuring the system performs at optimal levels under real-world workloads. \\
\paragraph{Resource Optimization} Optimize hardware and software resources to achieve better scalability and efficiency in post-quantum systems. This includes making decisions on hardware acceleration (for example, using specialized cryptographic processors, GPUs, or Field Programmable Gate Array (FPGAs) and optimizing memory usage and network bandwidth). Efficient resource allocation helps reduce costs and improve the overall responsiveness of cryptographic operations, especially as new algorithms are introduced or existing systems are modified. \\
\paragraph{Process Improvement} Refine internal processes for managing PQC systems, such as key management, certificate issuance, encryption/decryption workflows, and incident response. Use feedback from previous reviews, performance tests, and audits to implement process changes that improve system reliability, reduce complexity, and enhance agility. Automating key aspects of the cryptographic lifecycle, such as key rotation and certificate management, can reduce human error and improve operational efficiency. \\
\paragraph{Cost Reduction} Focus on cost reduction strategies by identifying areas where resources such as hardware, bandwidth, and power consumption can be used more effectively. Evaluate cloud or hybrid solutions for their potential to lower infrastructure costs, or consider using more efficient post-quantum algorithms that balance security and computational load. Analyzing cost-benefit tradeoffs between different cryptographic approaches (for example, RSA versus lattice-based algorithms) helps optimize the overall cost of deployment and operation.\\
\paragraph{Efficiency Enhancement} Enhance system efficiency by optimizing both the cryptographic algorithms and the underlying infrastructure. This includes improving the speed of key generation, reducing computational complexity, and utilizing modern cryptographic libraries that are optimized for post-quantum use cases. Efficiency enhancements help ensure that the organization can continue to meet user demands without overburdening computational resources, even as quantum-safe cryptographic algorithms become more complex.

\subsubsection{Evolution Planning}
Evolution Planning ensures that the organization's PQC systems remain adaptable to future developments in technology, standards, and threat landscapes. This stage involves proactively tracking changes in cryptographic research, monitoring industry standards, assessing emerging threats, enhancing capabilities, and making strategic updates to stay ahead of quantum risks and maintain a competitive edge.\\
Key elements of this phase include Technology tracking, Standards monitoring, Threat assessment, Capability enhancement, and Strategic updates.
\paragraph{Technology Tracking} Continuously monitor developments in quantum computing, cryptographic research, and related technologies to anticipate breakthroughs that could affect cryptographic security. Tracking advancements in quantum algorithms (such as Shor’s algorithm and Grover’s algorithm \cite{zhang2020depth}, quantum-safe cryptography, and PQC standards) will allow the organization to quickly adapt to new challenges and opportunities in the cryptographic domain.\\
\paragraph{Standards Monitoring} Stay up to date with emerging and evolving cryptographic standards, particularly those being developed by bodies such as NIST, ETSI, and ISO. This includes actively monitoring the progress of the NIST PQC Standardization project, as well as any updates to established protocols or guidelines that may impact post-quantum systems. Standards monitoring ensures that the organization’s cryptographic practices align with global best practices and regulatory expectations. \\
\paragraph{Threat Assessment} Conduct ongoing threat assessments to evaluate the impact of emerging quantum and classical attacks on the organization’s cryptographic infrastructure. This involves analyzing quantum capabilities, both at the academic and practical implementation levels, as well as assessing the evolution of classical cryptographic attacks. By regularly updating threat models, the organization can prioritize cryptographic solutions that mitigate new and emerging risks.\\
\paragraph{Capability Enhancement} Develop and enhance cryptographic capabilities in response to evolving business needs and technological advancements. This includes adopting new cryptographic algorithms as they become standardized, improving integration with quantum-safe protocols, and enhancing cryptographic efficiency. Capability enhancement also involves expanding the cryptographic skillset within the organization, enabling teams to respond more effectively to new challenges.\\
\paragraph{Strategic Updates} Make strategic updates to the organization's PQC roadmap based on insights gained from ongoing monitoring, threat assessments, and industry developments. These updates should align with the broader business strategy, ensuring that cryptographic transformations support the organization’s long-term objectives. Strategic updates may include re-evaluating the quantum readiness timeline, adjusting priorities, or implementing new technologies that align with future quantum and cryptographic trends.

\subsection{Stakeholder Management}
The Stakeholder Management phase is crucial for ensuring that all relevant parties within the organization are well-informed, engaged, and prepared for the changes brought by PQC \cite{radanliev2024ethics}. Effective internal communication ensures alignment, minimizes resistance, and fosters cooperation as cryptographic systems evolve.

\subsubsection{Internal Communication}
Effective Internal Communication is essential for ensuring smooth adoption of PQC systems. This involves keeping employees, technical teams, and business leaders informed about progress, changes, and the impact of the post-quantum transition.\\
This stage encompasses Progress reporting, Training programs, Change management, Impact notifications, and Support channels as its core components.
\paragraph{Progress Reporting} Establish regular communication channels to report on the status of the post-quantum transition. These reports should provide updates on milestones, accomplishments, timelines, and any challenges encountered during the implementation. Regular progress reports—whether through email updates, internal newsletters, or team meetings—ensure that stakeholders are aware of the ongoing work and the strategic importance of the transition. \\
\paragraph{Training Programs} Implement targeted training programs to ensure that employees, from developers to executive leadership, understand the implications of PQC. Training should cover cryptographic fundamentals, the specifics of quantum-safe algorithms, and the security policies related to cryptographic systems. Specialized training sessions should also be offered to key technical teams responsible for implementation, maintenance, and incident response. \\
\paragraph{Change Management} A structured change management process should be established to guide the organization through the shifts in cryptographic practices. This includes defining clear processes for implementing new technologies, shifting existing workflows, and addressing any concerns from impacted teams. Change management should focus on providing clear communication regarding the reasons for changes, the benefits of the new systems, and how employees will be supported during the transition. Creating a change network of champions across departments can facilitate smoother adoption. \\
\paragraph{Impact Notifications} Notify internal stakeholders about significant impacts related to the post-quantum transition. These notifications may include upcoming system downtimes, changes in user authentication methods, or updates to encryption standards that require operational adjustments. Notifications should be issued in advance, with clear explanations of the expected impact and actions required by the recipients. Timely and transparent communication helps minimize disruptions and prepares teams for changes in daily operations. \\
\paragraph{Support Channels} Provide clear and accessible support channels to assist employees with any issues or questions related to the post-quantum transition. This may include a dedicated helpdesk for cryptographic-related inquiries, self-service knowledge bases, Frequently Asked Questions (FAQs), and escalation procedures for technical issues. A responsive support system ensures that stakeholders can resolve challenges promptly, reducing frustration and facilitating smoother implementation.

\subsubsection{External Relations}
Effective management of external relations is critical for ensuring transparency, trust, and collaboration with customers, partners, vendors, regulators, and the broader public \cite{csenkey2023post}. As organizations transition to PQC, it’s essential to communicate changes clearly and maintain strong relationships with external stakeholders to support smooth operations and compliance. \\
This phase encompasses key efforts such as Customer communication, Partner coordination, Vendor management, Regulatory reporting, and Public Relations (PRs).
\paragraph{Customer Communication} Regular and transparent customer communication is vital to ensure that clients are aware of any changes that may affect their services or security. This includes informing customers about the adoption of PQC standards, how these changes enhance security, and any necessary steps they may need to take (such as software updates or changes in data handling procedures). Customer notifications should be clear, concise, and tailored to the level of technical detail appropriate for different customer segments. \\
\paragraph{Partner Coordination} Collaborate closely with external business partners to ensure that their systems, protocols, and cryptographic practices align with the organization’s post-quantum initiatives. This coordination may involve joint assessments of cryptographic standards, shared security practices, or ensuring mutual compatibility of encryption methods across platforms. Regular meetings, workshops, or updates can help maintain alignment and prevent security or integration issues.\\
\paragraph{Vendor Management} Manage relationships with vendors to ensure that third-party cryptographic solutions are compliant with post-quantum requirements. This includes negotiating contracts that outline vendor responsibilities for supporting quantum-safe algorithms, ensuring compatibility with the organization’s evolving cryptographic infrastructure, and setting clear expectations for software updates, patches, and support. Vendor risk assessments should be conducted regularly to evaluate the security and reliability of their cryptographic solutions. \\
\paragraph{Regulatory Reporting} Compliance with regulatory requirements is essential, especially in industries with strict data protection and cryptographic standards. Maintain clear communication with regulatory bodies to ensure the organization meets all reporting and auditing requirements related to PQC practices. This may involve submitting reports or certifications about the security posture, key management practices, or algorithmic updates. Staying ahead of regulatory changes will also ensure the organization remains compliant as new laws and guidelines are enacted in response to quantum threats. \\
\paragraph{Public Relations} Effective public relations strategies are important for maintaining a positive public image during the transition to PQC. This includes crafting messaging that highlights the organization’s commitment to advanced security measures and its proactive approach to addressing emerging quantum threats. PR efforts should focus on educating the public about the benefits of PQC and the organization’s leadership in securing data against future threats. Engaging with media, conducting webinars, and publishing white papers are effective ways to raise awareness and position the organization as a forward-thinking leader in cybersecurity.

\subsubsection{Documentation}
Comprehensive and well-maintained Documentation is essential for ensuring that the organization’s PQC transformation is well-understood, auditable, and compliant \cite{kannwischer2022improving}. Proper documentation serves as both a reference for internal stakeholders and an essential tool for ensuring ongoing maintenance, training, and regulatory adherence. \\
This section addresses key documentation areas, including Technical Documentation, Process Documentation, Training Materials, Compliance Records, and Audit Trails.
\paragraph{Technical Documentation} Create detailed technical documentation that covers the architecture, implementation, and operational aspects of PQC systems. This should include descriptions of algorithms in use (both classical and quantum-safe), key management protocols, system architecture diagrams, cryptographic libraries, integration guides, and security configurations. Technical documentation provides a foundational resource for developers, engineers, and security teams, enabling them to understand the system’s cryptographic framework, troubleshoot issues, and ensure future scalability and upgrades. \\
\paragraph{Process Documentation} Document the processes related to the adoption, implementation, and maintenance of PQC. This should cover operational workflows for cryptographic key generation and management, certificate issuance and lifecycle management, encryption protocols, and incident response procedures. Clear process documentation ensures consistency, supports knowledge transfer, and facilitates adherence to organizational and security best practices. It is also valuable during audits or when integrating new team members into the project. \\
\paragraph{Training Materials} Develop training materials to ensure that all relevant stakeholders are equipped with the necessary knowledge to work with post-quantum systems. These materials should cater to different skill levels, from introductory content for non-technical personnel to in-depth guides for cryptography experts. Training materials may include presentations, user manuals, video tutorials, and hands-on exercises. Regular updates to training content should be made as the post-quantum landscape evolves, ensuring that employees remain knowledgeable and capable of supporting new cryptographic technologies.\\
\paragraph{Compliance Records} Maintain compliance records to ensure adherence to relevant regulations and standards in the PQC environment. These records should document steps taken to meet compliance requirements, such as selecting specific cryptographic algorithms, performing risk assessments, or adopting quantum-safe practices. Compliance records also include certifications from third-party audits and any communications with regulatory bodies. These documents play a critical role in demonstrating that the organization’s cryptographic systems align with industry standards and legal obligations, and they may be necessary for future audits or certifications. \\
\paragraph{Audit Trails} Establish and maintain audit trails for all cryptographic activities, including key creation, updates, usage, and revocations. Audit trails provide a transparent and traceable record of all cryptographic events, making it easier to identify any anomalies, errors, or unauthorized actions. These records should be stored securely and be accessible for review by security teams, auditors, or compliance officers. Ensuring the integrity of audit trails is crucial for post-quantum systems, as they may be subject to more stringent security and compliance requirements over time.

\subsection{Success Metrics}
To measure the effectiveness and progress of the PQC transformation, it is essential to define clear success metrics \cite{fathalla2024beyond}. These metrics provide a way to gauge whether the project is achieving its objectives and meeting performance standards. Operational metrics focus on evaluating the ongoing health and efficiency of the system, its implementation, and its acceptance within the organization.

\subsubsection{Operational Metrics}
Operational metrics are essential for evaluating the effectiveness, efficiency, and impact of the PQC transition. These metrics provide quantifiable insights into system behavior, resource allocation, implementation status, issue management, and user engagement. By closely monitoring these factors, organizations can ensure that PQC integration meets performance expectations, remains on schedule, and aligns with strategic goals.\\
This section addresses key operational metrics, including System Performance, Implementation Progress, Resource Utilization, Issue Resolution, and User Adoption.
\paragraph{System Performance} Track and evaluate the performance of the cryptographic systems throughout the post-quantum transition. This includes monitoring the speed of encryption and decryption processes, system responsiveness, and throughput. KPIs should include transaction times, processing delays, and resource consumption for post-quantum algorithms, ensuring that the cryptographic systems do not cause unacceptable slowdowns or degrade the user experience. \\
\paragraph{Implementation Progress} Measure the progress of implementation against the planned roadmap and milestones. This includes tracking the completion of phases such as architecture development, testing, and deployment, as well as the timely rollout of post-quantum solutions across different systems and applications. Regular check-ins and progress reporting will ensure the initiative stays on schedule and that any delays or roadblocks are identified and addressed promptly. \\
\paragraph{Resource Utilization} Monitor resource utilization during both the transition to and operation of post-quantum systems. This involves assessing the hardware, software, and human resources being used to implement and maintain PQC systems. Efficient resource utilization is critical to avoiding cost overruns and ensuring the sustainability of the systems. Metrics might include CPU/GPU load during cryptographic operations, memory usage, bandwidth consumption, and staff hours allocated to the initiative. \\
\paragraph{Issue Resolution} Track the issue resolution process\footnote{Issue resolution refers to the process of identifying and addressing problems, mitigating risks, and making necessary decisions to maintain progress and stability.} to evaluate how quickly and effectively issues are identified, escalated, and resolved. This includes security incidents, performance bottlenecks, integration challenges, and compliance discrepancies. KPIs for issue resolution should focus on response times, Mean Time To Recovery (MTTR), and the number of unresolved issues over time. Effective issue resolution demonstrates the resilience and maturity of the PQC infrastructure. \\
\paragraph{User Adoption} Evaluate user adoption of PQC systems. This involves measuring the degree to which internal and external users (such as employees, customers, or partners) are transitioning to and utilizing the new quantum-safe systems. Adoption metrics could include the percentage of users actively using quantum-safe encryption, the level of engagement with new protocols, and feedback from user surveys. High adoption rates indicate a successful integration of post-quantum systems into daily operations.

\subsubsection{Security Metrics}
Security metrics are critical for assessing the effectiveness of PQC systems in protecting the organization’s data and infrastructure \cite{oliva2024cybersecurity}. These metrics provide insights into the resilience of the system against quantum and classical threats, the organization's ability to detect and respond to security incidents, and its overall compliance with relevant security standards. \\
This section focuses on key security metrics, including Vulnerability Assessment, Risk Reduction, Incident Metrics, Compliance Status, and Security Posture.
\paragraph{Vulnerability Assessment}  Regular vulnerability assessments help identify weaknesses in PQC implementations. This includes the continuous scanning for potential flaws in algorithms, key management procedures, cryptographic libraries, and overall system configurations. Vulnerability metrics track the number and severity of discovered vulnerabilities, the time taken to patch or mitigate them, and the effectiveness of corrective measures. Lower vulnerability scores indicate a more secure post-quantum infrastructure. \\
\paragraph{Risk Reduction} Measure the reduction in risk achieved through the adoption of PQC solutions. This involves assessing the organization's exposure to quantum threats compared to pre-transition risks, such as susceptibility to quantum-enabled cryptographic attacks (such as breaking RSA or ECC algorithms). Key indicators include a decrease in the potential impact of data breaches, an improved cryptographic defense posture, and an overall reduction in the likelihood of successful quantum-based or classical cryptographic attacks. \\
\paragraph{Incident Metrics} Track incident metrics related to the PQC environment, including the number of security breaches, data leaks, or unauthorized access events that occur post-deployment. Incident metrics should also capture how quickly security incidents are detected, contained, and resolved such as Mean Time To Detect (MTTD), and MTTR. These metrics help gauge the system's readiness to detect and respond to threats in real-time and evaluate how the cryptographic infrastructure handles emerging vulnerabilities or exploit attempts. \\
\paragraph{Compliance Status} Measure the compliance status of the organization’s cryptographic practices against industry regulations and standards. This includes tracking adherence to both quantum-safe cryptographic standards such as NIST’s PQC standards and broader cybersecurity frameworks such as GDPR, HIPAA, or PCI DSS. Compliance metrics might involve passing audits, maintaining certifications, and ensuring that all cryptographic systems are updated and in line with the latest regulatory guidelines. Achieving compliance demonstrates the organization’s commitment to security and its ability to protect sensitive data in the face of evolving threats. \\
\paragraph{Security Posture} Evaluate the overall security posture of the organization’s cryptographic systems. This includes monitoring security configurations, the implementation of best practices, the robustness of cryptographic key management, and the alignment of deployed systems with post-quantum security frameworks. A strong security posture indicates that the organization is well-prepared to defend against both quantum and classical threats, with established mechanisms for ongoing monitoring, threat detection, and proactive risk mitigation. Metrics related to security posture should also include the frequency of system updates, the effectiveness of penetration testing, and overall system robustness.

\subsubsection{Business Metrics}
Business Metrics are essential for measuring the broader impact of the PQC transformation on the organization's overall performance, strategic alignment, and stakeholder satisfaction \cite{purohit2024building}. These metrics help ensure that the transition not only improves security but also supports business objectives such as cost management, continuity, customer trust, and market positioning.\\
This section focuses on key business metrics, including Cost Management, Business Continuity, Customer Satisfaction, Partner Readiness, and Market Position.
\paragraph{Cost Management} Cost management metrics track the financial implications of adopting PQC, including the costs associated with implementation, training, system upgrades, and vendor contracts. These metrics help evaluate whether the transformation is being executed within the allocated budget, and they also provide insights into the cost-effectiveness of different solutions. Metrics such as Total Cost of Ownership (TCO)\footnote{Total Cost of Ownership (TCO) is used to assess both the direct and indirect expenses associated with acquiring and operating a product or service}, Return On Investment (ROI)\footnote{Return on Investment (ROI), or return on costs, is a financial ratio that compares the net income generated to the total amount invested.}, and budget variance can help identify areas where costs can be reduced or optimized while maintaining robust security.\\
\paragraph{Business Continuity} Measure the effectiveness of post-quantum systems in ensuring business continuity. This includes tracking how well the cryptographic infrastructure can support uninterrupted business operations during the transition. Key indicators include RTO, Recovery Point Objectives (RPO)\footnote{The Recovery Point Objective (RPO) is the maximum tolerable amount of data loss, measured in time, that an organization can accept. It helps determine if the backup frequency is adequate for disaster recovery.}, system uptime, and resilience against both quantum-enabled and classical cybersecurity threats. High business continuity scores indicate that the organization’s critical systems and data are secure and accessible, even in the face of emerging quantum threats or cyber incidents. \\
\paragraph{Customer Satisfaction} Track customer satisfaction metrics to gauge the impact of the post-quantum transition on customer trust and confidence. This includes measuring customer feedback on the security of services, transparency of communication, and any changes in service offerings related to PQC. Customer satisfaction surveys, Net Promoter Scores (NPS)\footnote{Net Promoter Score (NPS) is a metric based on how likely customers are to recommend a company, product, or service to others.}, and Customer Retention Rates (CRRs)\footnote{Customer Retention Rate (CRR) measures the percentage of customers who continue to do business with a company over a specified time period.} can provide insights into how the post-quantum transition is perceived by clients and whether it enhances their confidence in the organization's ability to protect their data. Increased customer satisfaction signifies that the transition is improving the company's reputation as a leader in secure, future-proof services. \\
\paragraph{Partner Readiness} Monitor partner readiness metrics to ensure that external business partners, including suppliers, collaborators, and third-party service providers, are aligned with the organization’s PQC standards. This includes evaluating the extent to which partners have adopted quantum-safe cryptographic measures and ensuring that their systems are compatible with the organization’s infrastructure. Key metrics could include the percentage of partners that have completed post-quantum readiness assessments, the completion rate of contractual obligations related to cryptography, and the success of joint security initiatives. High partner readiness ensures seamless integration and reduces risk across the supply chain. \\
\paragraph{Market Position} Measure the organization’s market position relative to competitors in terms of adopting PQC and being a leader in cybersecurity innovation. Metrics such as market share, brand perception, thought leadership such as published white papers, conference participation, and media coverage on quantum security can indicate how well the company is positioned as a trusted leader in secure data management. A strong market position driven by cutting-edge post-quantum security can attract customers, partners, and talent, providing a competitive advantage in the marketplace.

\section{Quantum-Ready Architecture for Security and Risk Management (QUASAR) Model Framework}
The proposed framework presents a comprehensive model for preparing organizations for the post-quantum era. It encompasses strategic, technical, and operational components necessary for successful quantum-safe transition.
In addition, it provides a comprehensive approach for organizational quantum preparedness, incorporating technical, security, and operational components with clear metrics for success measurement and continuous improvement. Prior to discussing the model in detail, it is important to note that the proposed formulas serve merely as illustrative examples or baseline references, and other approaches may equally be applicable.

\subsection{Core Components}
At the heart of the QUASAR framework lies a composite model that quantifies an organization's overall readiness for the post-quantum era. This model is constructed around three foundational domains: Technical, Security, and Operational readiness. Each domain is evaluated independently and then integrated into a single readiness score Post-Quantum Readiness (PQR) through a weighted aggregation.
The model is formally defined by the following equation:
\begin{equation}
    PQR = \sum_{i=1}^{n} (T_i \cdot w_i + S_i \cdot w_i + O_i \cdot w_i)
\end{equation}

Subject to the normalization constraint:

\begin{equation}
    \sum_{i=1}^{n} w_i = 1
\end{equation}

Where:
\begin{itemize}
    \item $PQR$ = Overall Post-Quantum Readiness score, representing the organization's composite preparedness level.
    \item $T_i$ = Score for the $i^{th}$ component of Technical Readiness, including cryptographic systems, algorithm adoption, and infrastructure agility.
    \item $S_i$ = Score for the $i^{th}$ component of Security Readiness, such as key management, data protection policies, and incident response capabilities in a quantum context.
    \item $O_i$ = Score for the $i^{th}$ component of Operational Readiness, covering areas like workforce training, change management, vendor alignment, and governance.
    \item $w_i$ = Weight assigned to the $i^{th}$ component, reflecting its relative importance to the organization’s overall strategy.
\end{itemize}
This model enables organizations to customize the relative emphasis placed on different components based on their specific risk profile, industry requirements, or strategic objectives. By requiring the weights \( w_i \) to sum to 1, the model ensures balanced aggregation, allowing it to function as a normalized metric suitable for internal benchmarking, temporal comparison, or cross-industry analysis.\\
Furthermore, the modular nature of this approach supports flexibility and extensibility. As new quantum threats or technologies emerge, additional components can be incorporated into the existing structure without disrupting the overall model.

\subsection{Component Matrices}
As mentioned above, in the QUASAR framework, organizational readiness for the post-quantum era is assessed through three distinct but interrelated matrices: technical, security, and operational. Each matrix captures the specific dimensions relevant to its domain, providing a structured approach for evaluation and prioritization. These matrices help quantify readiness levels, identify gaps, and allocate resources effectively. We begin by detailing the technical readiness matrix, which focuses on the technological underpinnings of an organization’s quantum preparedness.

\subsubsection{Technical Readiness Matrix}
The Technical Readiness Matrix $T$ is designed to assess the organization’s current technical landscape in relation to post-quantum cybersecurity. It evaluates three core components across multiple domains: Cryptographic systems, Infrastructure dependencies, and Algorithm preparedness.

\begin{equation}
    T = \begin{bmatrix} 
    C_{11} & C_{12} & C_{13} \\
    I_{21} & I_{22} & I_{23} \\
    A_{31} & A_{32} & A_{33}
    \end{bmatrix}
\end{equation}

Where: \begin{itemize} \item $C_{ij}$ = Cryptographic Components — assessment of existing encryption schemes, key exchange protocols, and their vulnerability to quantum attacks. This includes both in-use cryptographic assets and legacy systems. \item $I_{ij}$ = Infrastructure Components — evaluation of system dependencies such as network architecture, hardware compatibility, and support for cryptographic agility (such as  support for post-quantum algorithms). \item $A_{ij}$ = Algorithm Components — preparedness of the organization to adopt, test, and deploy quantum-resistant algorithms. This includes participation in standardization efforts (such as NIST PQC), pilot deployments, and algorithmic diversity. \end{itemize}

Each element $T_{ij}$  within the matrix is scored or evaluated based on defined readiness criteria, allowing the organization to compute an overall technical readiness score or identify specific areas requiring attention. This matrix feeds into the broader PQR score, contributing alongside the Security and Operational matrices.

\section{Implementation Framework}
The QUASAR framework provides not only a conceptual foundation but also a structured path toward implementation. This section introduces two core components of the implementation phase: the phase transformation model, which captures the dynamic evolution of post-quantum readiness over time, and the risk assessment matrix, which aids in prioritizing mitigation efforts based on quantified risk factors.

\subsection{Phase Transformation}
The transition to post-quantum readiness is not instantaneous; it occurs gradually as organizations adapt their systems, policies, and technologies. We model this transformation over time using an exponential transition function:

\begin{equation}
    P(t) = \alpha \cdot e^{-\lambda t} + \beta \cdot (1 - e^{-\lambda t})
\end{equation}
Where:
\begin{itemize}
    \item $P(t)$ = The organization's preparedness level at time $t$
    \item $\alpha$ = The initial state of preparedness (at $t = 0$)
    \item $\beta$ = The desired or target state of preparedness
    \item $\lambda$ = The transformation rate (i.e., how quickly readiness evolves)
\end{itemize}

This equation models preparedness as an exponential decay from the initial state $\alpha$ toward the target state $\beta$. A higher value of $\lambda$ indicates a faster transformation, suggesting greater resource allocation, urgency, or organizational agility. This model can be used to forecast readiness under different strategic scenarios and resource investments.

\subsection{Risk Assessment Matrix}
To effectively prioritize actions within the QUASAR framework, we introduce a weighted Risk Assessment Matrix. This matrix quantifies different risk categories across various dimensions and combines them with corresponding weights to produce an aggregated risk score:
\begin{equation}
    R = \begin{bmatrix}
    r_{11} & r_{12} & r_{13} \\
    r_{21} & r_{22} & r_{23} \\
    r_{31} & r_{32} & r_{33}
    \end{bmatrix} \cdot \begin{bmatrix}
    w_1 \\
    w_2 \\
    w_3
    \end{bmatrix}
\end{equation}
Where:
\begin{itemize}
    \item $r_{ij}$ = Risk score of the $i^{th}$ category under the $j^{th}$ dimension (such as technical, operational, and security risks)
    \item $w_j$ = Weight assigned to the $j^{th}$ dimension, reflecting its relative importance
    \item $R$ = The resulting risk vector or score, which can be used to rank or prioritize mitigation efforts
    \item $\sum_{j=1}^{3} w_j = 1$ = The weights are subject to the constraint that their total must sum to 1
\end{itemize}
Each row of the matrix corresponds to a different risk category (such as cryptographic vulnerability, infrastructure exposure, or compliance risk), and each column corresponds to a dimension of readiness. The resulting product provides a weighted view of the organization's risk exposure, guiding strategic prioritization in resource allocation and mitigation planning.

\section{Implementation Phases}
The transition toward PQR within the QUASAR framework is structured into distinct phases to provide a logical, measurable progression. Each phase incorporates both qualitative and quantitative elements, allowing for dynamic assessment, transformation, and long-term integration.

\subsection{Phase I: Assessment}
The first phase involves understanding the organization's current posture, identifying gaps, and establishing a baseline from which transformation can occur.
\begin{enumerate}[label=\arabic*.]
\item \textbf{Initial Assessment}
The initial assessment aggregates multiple criteria that reflect the organization's preparedness across technical, security, and operational domains. Each criterion is weighted to reflect its relative importance.
\begin{equation}
A_i = \sum_{j=1}^{m} c_j \cdot w_j
\end{equation}
Where:
\begin{itemize}
\item $A_i$ = Aggregate assessment score for the $i^{th}$ area or domain
\item $c_j$ = Score or status of the $j^{th}$ individual criterion
\item $w_j$ = Weight of the $j^{th}$ criterion (with $\sum w_j = 1$)
\end{itemize}
This weighted sum allows organizations to tailor the assessment to their specific context by adjusting the weights according to strategic priorities or industry regulations.

\item \textbf{Gap Analysis}
The difference between the current and desired states is quantified as the gap value $G$, which helps prioritize areas for transformation.
\begin{equation}
G = T_d - C_s
\end{equation}
Where:
\begin{itemize}
\item $T_d$ = Target state value (such as desired security maturity level)
\item $C_s$ = Current state value (measured via assessment)
\end{itemize}
A larger $G$ indicates a greater deviation from the desired state, signaling the need for more focused intervention. This analysis drives resource allocation and strategic planning in subsequent phases.
\end{enumerate}

\subsection{Phase II: Implementation}
The second phase operationalizes the transformation strategy, executing projects, deploying new technologies, and updating protocols to move toward PQR. The process is modeled as an exponential function that captures the rate of implementation over time:
\begin{equation}
    I(t) = I_0 + (I_f - I_0)(1 - e^{-kt})
\end{equation}
Where:
\begin{itemize}
    \item $I(t)$ = Implementation progress at time $t$
    \item $I_0$ = Initial implementation state (such as current system adoption level)
    \item $I_f$ = Final or target implementation state (full deployment/readiness)
    \item $k$ = Implementation rate constant, influenced by resource availability, organizational readiness, and external constraints
\end{itemize}
This model illustrates diminishing returns over time, where early gains are more rapid and subsequent progress requires increasing effort. It supports project planning, timeline forecasting, and performance monitoring, aligning implementation milestones with strategic objectives.

\section{Success Metrics}
To evaluate the effectiveness of the QUASAR framework and its implementation over time, it is essential to define quantitative success metrics. These metrics provide measurable indicators of progress and allow stakeholders to track improvements, identify deficiencies, and benchmark organizational readiness against post-quantum security standards.

\subsection{Performance Indicators}
Performance Indicators (PI) are designed to capture the overall effectiveness of the implemented measures across various domains such as technical readiness, operational efficiency, compliance, and risk mitigation. Each metric is assigned a weight to reflect its strategic importance.
\begin{equation}
    PI = \frac{1}{n} \sum_{i=1}^{n} (w_i \cdot M_i)
\end{equation}
Where:
\begin{itemize}
\item $PI$ = Composite Performance Indicator
    \item $M_i$ = Score of the $i^{\text{th}}$ individual metric (such as encryption upgrade completion, policy compliance)
    \item $w_i$ = Weight assigned to the $i^{\text{th}}$ metric, where $\sum_{i=1}^{n} w_i = 1$
    \item $n$ = Total number of metrics considered
\end{itemize}
This weighted average allows for normalization and comparability across organizations of different sizes or industries. The use of weights also ensures that more critical metrics have a proportionally larger impact on the overall performance evaluation.

\subsection{Readiness Score}
The Readiness Score (RS) offers a holistic measure of an organization's current state of preparedness for the post-quantum era. Unlike the average-based PI, this score uses a Root Sum Square (RSS) approach to emphasize higher-magnitude gaps or risks.
\begin{equation}
    RS = \sqrt{\sum_{i=1}^{n} (w_i \cdot r_i)^2}
\end{equation}
Where:
\begin{itemize}
    \item $RS$ = Overall Readiness Score
    \item $r_i$ = Readiness level or maturity score for the $i^{th}$ component
    \item $w_i$ = Weight of the $i^{th}$ component (reflecting strategic importance), where $\sum_{i=1}^{n} w_i = 1$
    \item $n$ = Total number of readiness components evaluated
\end{itemize}
The squaring operation in this model penalizes lower-performing components more heavily, highlighting areas of concern that may otherwise be masked in an average-based system. A higher RS value indicates a stronger, more balanced readiness profile, while a lower score suggests a need for targeted intervention.\\
These metrics, when tracked over time, support continuous improvement, progress reporting, and stakeholder communication regarding the organization’s journey toward PQR.

\section{Timeline Framework}
Effective implementation of a post-quantum readiness strategy requires careful temporal planning. The QUASAR framework divides the timeline into three strategic horizons: Short-term, Medium-term, and Long-term actions. Each horizon is governed by a time-dependent function that reflects how the urgency, impact, and scope of activities evolve over time. These functions incorporate an exponential factor to model diminishing urgency for immediate actions and increasing relevance for future efforts.

\subsection{Short-term Actions}
Short-term actions are immediate steps that must be prioritized to build the foundation for long-term transformation. These include activities such as asset inventory, cryptographic audits, risk assessments, and stakeholder engagement. The urgency of these actions decays exponentially over time as their relevance diminishes once addressed.

\begin{equation}
    ST(t) = \sum_{i=1}^{n} A_i(t) \cdot e^{-\lambda t}
\end{equation}
Where:
\begin{itemize}
    \item $ST(t)$ = Aggregate value of short-term action impact at time $t$
    \item $A_i(t)$ = Time-dependent contribution of the $i^{th}$ short-term action
    \item $\lambda$ = Decay rate constant (higher values imply faster urgency decay)
    \item $t$ = Time since initiation of the framework
\end{itemize}
This formulation ensures that early actions receive the most attention and are not deferred inappropriately, reinforcing the principle of "early wins."

\subsection{Medium-term Actions}
Medium-term actions focus on transitioning core infrastructure and processes. These may include integrating quantum-resistant algorithms, updating vendor contracts, or redesigning internal protocols. Their importance rises over time as the organization moves beyond foundational steps.
\begin{equation}
    MT(t) = \sum_{i=1}^{n} A_i(t) \cdot (1 - e^{-\lambda t})
\end{equation}
Where:
\begin{itemize}
    \item $MT(t)$ = Aggregate value of medium-term action impact at time $t$
    \item $(1 - e^{-\lambda t})$ = Growth function indicating increasing importance over time
\end{itemize}
This exponential growth curve models a smooth ramp-up in implementation activities as short-term goals are completed and organizational capacity increases.

\subsection{Long-term Actions}
Long-term actions address strategic resilience and adaptive capacity. These include continuous monitoring, policy revisions, advanced threat modeling, and participation in post-quantum research or standardization efforts. Their significance emerges more prominently in the later stages of the implementation lifecycle.
\begin{equation}
    LT(t) = \sum_{i=1}^{n} A_i(t) \cdot (1 - e^{-2\lambda t})
\end{equation}
The factor $2\lambda$ ensures that long-term actions ramp up more slowly initially but become dominant in the later stages. This formulation reflects the reality that long-term resilience requires sustained, forward-looking investment beyond initial implementation milestones.

\section{Continuous Improvement Metrics}
Long-term success in PQR depends not only on initial implementation but also on sustained optimization. The QUASAR framework supports a continuous improvement cycle through strategic monitoring, performance feedback, and adaptive resource allocation. This is formalized via an optimization function that seeks to improve key objectives while operating under relevant constraints such as resource availability, regulatory limits, and technical feasibility.

\subsection{Optimization Function}
The goal of continuous improvement is to maximize overall organizational benefit or efficiency, as expressed by an objective function composed of several contributing factors.
\begin{equation}
    O(t) = \max_{x \in X} \sum_{i=1}^{n} f_i(x,t)
\end{equation}
Where:
\begin{itemize}
    \item $O(t)$ = Optimal value of the organization’s performance or readiness at time $t$
    \item $x \in X$ = Set of decision variables (such as budget allocations, personnel assignments, tool selections)
    \item $f_i(x,t)$ = Time-dependent objective function components (such as minimized risk, maximized cryptographic coverage, improved training effectiveness)
    \item $n$ = Number of objectives or contributing factors being optimized
\end{itemize}
This formulation allows for multiple performance dimensions to be jointly considered, enabling trade-offs and prioritization based on organizational strategy and external changes.
The optimization is subject to constraints representing real-world limitations (subject to):
\begin{equation}
    \begin{cases}
        g_j(x,t) \leq 0, & j = 1,\ldots,m \\
        h_k(x,t) = 0, & k = 1,\ldots,p
    \end{cases}
\end{equation}
Where:
\begin{itemize}
    \item $g_j(x,t)$ = Inequality constraints (such as cost limits, regulatory thresholds, operational capacity)
    \item $h_k(x,t)$ = Equality constraints (such as balance between departments, system compatibility requirements)
    \item $m$ = Number of inequality constraints
    \item $p$ = Number of equality constraints
\end{itemize}
These constraints ensure that optimization remains grounded in operational realities while seeking the best possible outcome over time. This approach supports continuous recalibration, allowing organizations to respond to new threats, standards, or opportunities as the post-quantum landscape evolves.

\section{Discussion}
The QUASAR framework presents a structured and scalable approach to addressing the cybersecurity challenges posed by quantum computing. By breaking down readiness into three critical domains—technical, security, and operational—it offers a comprehensive pathway for organizations to prepare for the post-quantum era. However, the implementation and adoption of QUASAR are not without challenges, and several considerations must be taken into account.

\subsection{Challenges in Implementation}
One of the primary challenges organizations will face in implementing QUASAR is the complexity of transitioning existing systems. While the framework provides a roadmap for organizations to follow, the practical aspects of implementing quantum-resistant algorithms, updating infrastructure, and retraining staff can be resource-intensive. Many organizations may struggle to balance the urgency of short-term actions with the long-term commitment required for quantum readiness. Additionally, achieving alignment between stakeholders across departments, each with different priorities, can complicate the execution of a cohesive post-quantum strategy.\\
Moreover, the rapidly evolving nature of quantum computing introduces an element of uncertainty. While quantum computing research is progressing at a fast pace, predicting the exact timeline for the widespread availability of quantum computers capable of breaking current encryption methods remains challenging. This uncertainty can make it difficult for organizations to prioritize quantum readiness initiatives and allocate resources effectively. As a result, organizations may need to adopt a flexible approach that can quickly adapt to new developments in the field.

\subsection{Addressing the Limitations}
Despite these challenges, the QUASAR framework provides flexibility through its modular design. The ability to tailor the technical, security, and operational readiness components ensures that organizations of different sizes, industries, and maturity levels can adopt the framework at their own pace. The inclusion of optimization functions and continuous improvement cycles also ensures that readiness can be adjusted in real-time, based on emerging threats and the evolving quantum landscape.\\
Furthermore, the use of performance indicators and readiness scores offers a quantifiable method for tracking progress. These metrics provide organizations with clear benchmarks to measure their success in achieving post-quantum resilience. The optimization function’s consideration of constraints—whether technical, financial, or regulatory—also helps organizations identify and mitigate potential risks early in the process, ensuring that they stay aligned with both internal objectives and external requirements.

\subsection{Future Directions}
Looking ahead, the QUASAR framework can be further refined and expanded to account for new developments in the quantum computing space. For instance, as quantum algorithms and computational models evolve, organizations may need to adapt their cryptographic systems beyond current post-quantum algorithms. The continuous improvement model embedded in QUASAR is critical for this, as it allows organizations to dynamically respond to these advancements, integrating new cryptographic techniques as they become available.\\
Additionally, as more industry-specific guidelines and standards for quantum readiness are developed, the QUASAR framework could be enhanced to align more closely with these emerging norms. This would allow the framework to provide more targeted advice and strategies for sectors such as finance, healthcare, and government, each of which has unique security requirements and regulatory constraints.\\
Another promising area for future research lies in the integration of AI with the QUASAR framework. AI-based solutions could assist in automating parts of the readiness process, such as risk assessments, vulnerability scans, and threat modeling, to provide more precise and real-time evaluations of an organization's readiness.

\section{Conclusion}
The transition to a post-quantum world presents both unprecedented challenges and critical opportunities for organizations across all sectors. As quantum computing technologies mature, the vulnerabilities of classical cryptographic systems become increasingly urgent, demanding strategic foresight and structured response. The proposed QUASAR framework offers a comprehensive approach for navigating this complex landscape.\\
Through the integration of component-based readiness matrices, phase-wise implementation models, and mathematically grounded performance and optimization functions, QUASAR provides a holistic pathway for organizations to evaluate, act, and adapt. Each phase—assessment, implementation, and continuous improvement—is supported by quantifiable metrics and decision models that allow organizations to monitor progress and reallocate resources dynamically.\\
The framework emphasizes the importance of tailored strategies across technical, security, and operational domains, ensuring that post-quantum readiness is not treated as a one-time upgrade, but as a continuous, evolving discipline. The inclusion of a timeline model and success metrics further ensures that efforts are not only measurable but aligned with long-term business and security objectives.\\
In conclusion, preparing for the post-quantum era is no longer optional, it is a strategic imperative. QUASAR empowers organizations with the tools, models, and methodology required to future-proof their operations, protect sensitive assets, and remain resilient in the face of emerging quantum-era threats.


\bibliographystyle{IEEEtran}
\bibliography{ref.bib}

\begin{thebibliography}{10}
\providecommand{\url}[1]{#1}
\csname url@samestyle\endcsname
\providecommand{\newblock}{\relax}
\providecommand{\bibinfo}[2]{#2}
\providecommand{\BIBentrySTDinterwordspacing}{\spaceskip=0pt\relax}
\providecommand{\BIBentryALTinterwordstretchfactor}{4}
\providecommand{\BIBentryALTinterwordspacing}{\spaceskip=\fontdimen2\font plus
\BIBentryALTinterwordstretchfactor\fontdimen3\font minus \fontdimen4\font\relax}
\providecommand{\BIBforeignlanguage}[2]{{%
\expandafter\ifx\csname l@#1\endcsname\relax
\typeout{** WARNING: IEEEtran.bst: No hyphenation pattern has been}%
\typeout{** loaded for the language `#1'. Using the pattern for}%
\typeout{** the default language instead.}%
\else
\language=\csname l@#1\endcsname
\fi
#2}}
\providecommand{\BIBdecl}{\relax}
\BIBdecl

\bibitem{mavroeidis2018impact}
V.~Mavroeidis, K.~Vishi, M.~D. Zych, and A.~J{\o}sang, ``The impact of quantum computing on present cryptography,'' \emph{arXiv preprint arXiv:1804.00200}, 2018.

\bibitem{ajala2024exploring}
O.~A. Ajala, C.~A. Arinze, O.~C. Ofodile, C.~C. Okoye, and A.~I. Daraojimba, ``Exploring and reviewing the potential of quantum computing in enhancing cybersecurity encryption methods,'' \emph{Magna Sci. Adv. Res. Rev}, vol.~10, no.~1, pp. 321--329, 2024.

\bibitem{baseri2024cybersecurity}
Y.~Baseri, V.~Chouhan, and A.~Ghorbani, ``Cybersecurity in the quantum era: Assessing the impact of quantum computing on infrastructure,'' \emph{arXiv preprint arXiv:2404.10659}, 2024.

\bibitem{geremew2024preparing}
A.~Geremew and A.~Mohammad, ``Preparing critical infrastructure for post-quantum cryptography: Strategies for transitioning ahead of cryptanalytically relevant quantum computing,'' \emph{International Journal on Engineering, Science and Technology}, vol.~6, no.~4, pp. 338--365, 2024.

\bibitem{aydeger2024towards}
A.~Aydeger, E.~Zeydan, A.~K. Yadav, K.~T. Hemachandra, and M.~Liyanage, ``Towards a quantum-resilient future: Strategies for transitioning to post-quantum cryptography,'' in \emph{2024 15th International Conference on Network of the Future (NoF)}.\hskip 1em plus 0.5em minus 0.4em\relax IEEE, 2024, pp. 195--203.

\bibitem{hasan2024framework}
K.~F. Hasan, L.~Simpson, M.~A.~R. Baee, C.~Islam, Z.~Rahman, W.~Armstrong, P.~Gauravaram, and M.~McKague, ``A framework for migrating to post-quantum cryptography: Security dependency analysis and case studies,'' \emph{IEEE Access}, vol.~12, pp. 23\,427--23\,450, 2024.

\bibitem{nather2024migrating}
C.~N{\"a}ther, D.~Herzinger, S.-L. Gazdag, J.-P. Stegh{\"o}fer, S.~Daum, and D.~Loebenberger, ``Migrating software systems towards post-quantum cryptography--a systematic literature review,'' \emph{IEEE Access}, 2024.

\bibitem{joseph2022transitioning}
D.~Joseph, R.~Misoczki, M.~Manzano, J.~Tricot, F.~D. Pinuaga, O.~Lacombe, S.~Leichenauer, J.~Hidary, P.~Venables, and R.~Hansen, ``Transitioning organizations to post-quantum cryptography,'' \emph{Nature}, vol. 605, no. 7909, pp. 237--243, 2022.

\bibitem{fathalla2024beyond}
E.~Fathalla and M.~Azab, ``Beyond classical cryptography: A systematic review of post-quantum hash-based signature schemes, security, and optimizations,'' \emph{IEEE Access}, 2024.

\bibitem{grote2019review}
O.~Grote, A.~Ahrens, and C.~Benavente-Peces, ``A review of post-quantum cryptography and crypto-agility strategies,'' in \emph{2019 International Interdisciplinary PhD Workshop (IIPhDW)}.\hskip 1em plus 0.5em minus 0.4em\relax IEEE, 2019, pp. 115--120.

\bibitem{bishwas2024strategic}
A.~K. Bishwas and M.~Sen, ``Strategic roadmap for quantum-resistant security: A framework for preparing industries for the quantum threat,'' \emph{arXiv preprint arXiv:2411.09995}, 2024.

\bibitem{kappler2022post}
S.~A. K{\"a}ppler and B.~Schneider, ``Post-quantum cryptography: An introductory overview and implementation challenges of quantum-resistant algorithms,'' \emph{Proceedings of the Society}, vol.~84, pp. 61--71, 2022.

\bibitem{joshi2024guarding}
A.~Joshi, P.~Bhalgat, P.~Chavan, T.~Chaudhari, and S.~Patil, ``Guarding against quantum threats: A survey of post-quantum cryptography standardization, techniques, and current implementations,'' in \emph{International Conference on Applications and Techniques in Information Security}.\hskip 1em plus 0.5em minus 0.4em\relax Springer, 2024, pp. 33--46.

\bibitem{demir2025performance}
E.~D. Demir, B.~Bilgin, and M.~C. Onbasli, ``Performance analysis and industry deployment of post-quantum cryptography algorithms,'' \emph{arXiv preprint arXiv:2503.12952}, 2025.

\bibitem{kong2024realizing}
I.~Kong, M.~Janssen, and N.~Bharosa, ``Realizing quantum-safe information sharing: Implementation and adoption challenges and policy recommendations for quantum-safe transitions,'' \emph{Government Information Quarterly}, vol.~41, no.~1, p. 101884, 2024.

\bibitem{marchesi2025survey}
L.~Marchesi, M.~Marchesi, and R.~Tonelli, ``A survey on cryptoagility and agile practices in the light of quantum resistance,'' \emph{Information and Software Technology}, vol. 178, p. 107604, 2025.

\bibitem{okika2025assessing}
N.~Okika, G.~A. Nwatuzie, H.~S. Olarinoye, A.~A. Nwaka, E.~Igba, and R.~Dunee, ``Assessing the vulnerability of traditional and post-quantum cryptographic systems through penetration testing and strengthening cyber defenses with zero trust security in the era of quantum computing,'' \emph{International Journal of Innovative Science and Research Technology}, vol.~10, no.~2, 2025.

\bibitem{ricci2024hybrid}
S.~Ricci, P.~Dobias, L.~Malina, J.~Hajny, and P.~Jedlicka, ``Hybrid keys in practice: combining classical, quantum and post-quantum cryptography,'' \emph{IEEE Access}, vol.~12, pp. 23\,206--23\,219, 2024.

\bibitem{dasquantum}
P.~Das, ``Quantum computing in payments security: Preparing for the post-quantum era.''

\bibitem{csenkey2023post}
K.~Csenkey and N.~Bindel, ``Post-quantum cryptographic assemblages and the governance of the quantum threat,'' \emph{Journal of Cybersecurity}, vol.~9, no.~1, p. tyad001, 2023.

\bibitem{rayhan2024cybersecurity}
A.~Rayhan, ``Cybersecurity in the digital age: Assessing threats and strengthening defenses,'' in \emph{Conference: Cybersecurity Awareness}, 2024, pp. 1--26.

\bibitem{joshi2024emerging}
H.~Joshi, ``Emerging technologies driving zero trust maturity across industries,'' \emph{IEEE Open Journal of the Computer Society}, 2024.

\bibitem{dvorak2024leveraging}
R.~Dvorak and F.~Bahadori, ``Leveraging open source tools to teach quantum computing foundations: Bridging the future workforce gap in the quantum era,'' in \emph{2024 ASEE Annual Conference \& Exposition}, 2024.

\bibitem{marmebro2024investigation}
A.~Marmebro and K.~Stenbom, ``Investigation of post-quantum cryptography (fips 203 \& 204) compared to legacy cryptosystems, and implementation in large corporations.'' 2024.

\bibitem{hughes2022assessing}
C.~Hughes, D.~Finke, D.-A. German, C.~Merzbacher, P.~M. Vora, and H.~Lewandowski, ``Assessing the needs of the quantum industry,'' \emph{IEEE Transactions on Education}, vol.~65, no.~4, pp. 592--601, 2022.

\bibitem{meshram2015efficient}
C.~Meshram, ``An efficient id-based cryptographic encryption based on discrete logarithm problem and integer factorization problem,'' \emph{Information Processing Letters}, vol. 115, no.~2, pp. 351--358, 2015.

\bibitem{guneysu2012practical}
T.~G{\"u}neysu, V.~Lyubashevsky, and T.~P{\"o}ppelmann, ``Practical lattice-based cryptography: A signature scheme for embedded systems,'' in \emph{Cryptographic Hardware and Embedded Systems--CHES 2012: 14th International Workshop, Leuven, Belgium, September 9-12, 2012. Proceedings 14}.\hskip 1em plus 0.5em minus 0.4em\relax Springer, 2012, pp. 530--547.

\bibitem{impagliazzo1995personal}
R.~Impagliazzo, ``A personal view of average-case complexity,'' in \emph{Proceedings of Structure in Complexity Theory. Tenth Annual IEEE Conference}.\hskip 1em plus 0.5em minus 0.4em\relax IEEE, 1995, pp. 134--147.

\bibitem{ugwuishiwu2020overview}
C.~Ugwuishiwu, U.~Orji, C.~Ugwu, and C.~Asogwa, ``An overview of quantum cryptography and shor’s algorithm,'' \emph{Int. J. Adv. Trends Comput. Sci. Eng}, vol.~9, no.~5, 2020.

\bibitem{morimae2022quantum}
T.~Morimae and T.~Yamakawa, ``Quantum commitments and signatures without one-way functions,'' in \emph{Annual International Cryptology Conference}.\hskip 1em plus 0.5em minus 0.4em\relax Springer, 2022, pp. 269--295.

\bibitem{weinberg2025dynamic}
A.~I. Weinberg, ``Dynamic data defense: Unveiling the database in motion chaos encryption (dache) algorithm--a breakthrough in chaos theory for enhanced database security,'' \emph{arXiv preprint arXiv:2501.03296}, 2025.

\bibitem{bos2018crystals}
J.~Bos, L.~Ducas, E.~Kiltz, T.~Lepoint, V.~Lyubashevsky, J.~M. Schanck, P.~Schwabe, G.~Seiler, and D.~Stehl{\'e}, ``Crystals-kyber: a cca-secure module-lattice-based kem,'' in \emph{2018 IEEE European Symposium on Security and Privacy (EuroS\&P)}.\hskip 1em plus 0.5em minus 0.4em\relax IEEE, 2018, pp. 353--367.

\bibitem{ajtai1996generating}
M.~Ajtai, ``Generating hard instances of lattice problems,'' in \emph{Proceedings of the twenty-eighth annual ACM symposium on Theory of computing}, 1996, pp. 99--108.

\bibitem{lyubashevsky2020crystals}
V.~Lyubashevsky, L.~Ducas, E.~Kiltz, T.~Lepoint, P.~Schwabe, G.~Seiler, D.~Stehl{\'e}, and S.~Bai, ``Crystals-dilithium,'' \emph{Algorithm Specifications and Supporting Documentation}, 2020.

\bibitem{mohan2023hash}
P.~A. Mohan, ``Hash-based digital signatures-a tutorial review,'' in \emph{2023 IEEE International Conference on Public Key Infrastructure and its Applications (PKIA)}.\hskip 1em plus 0.5em minus 0.4em\relax IEEE, 2023, pp. 1--8.

\bibitem{saarinen2024accelerating}
M.-J.~O. Saarinen, ``Accelerating slh-dsa by two orders of magnitude with a single hash unit,'' in \emph{Annual International Cryptology Conference}.\hskip 1em plus 0.5em minus 0.4em\relax Springer, 2024, pp. 276--304.

\bibitem{hulsing2018rfc}
A.~H{\"u}lsing, D.~Butin, S.~Gazdag, J.~Rijneveld, and A.~Mohaisen, ``Rfc 8391: Xmss: extended merkle signature scheme,'' 2018.

\bibitem{dey2023progress}
J.~Dey and R.~Dutta, ``Progress in multivariate cryptography: Systematic review, challenges, and research directions,'' \emph{ACM Computing Surveys}, vol.~55, no.~12, pp. 1--34, 2023.

\bibitem{balamurugan2021post}
C.~Balamurugan, K.~Singh, G.~Ganesan, and M.~Rajarajan, ``Post-quantum and code-based cryptography—some prospective research directions,'' \emph{Cryptography}, vol.~5, no.~4, p.~38, 2021.

\bibitem{kuznetsov2021performance}
A.~Kuznetsov, M.~Lutsenko, M.~Bagmut, and V.~Zhora, ``Performance evaluation of the classic mceliece key encapsulation algorithm,'' in \emph{2021 11th IEEE International Conference on Intelligent Data Acquisition and Advanced Computing Systems: Technology and Applications (IDAACS)}, vol.~2.\hskip 1em plus 0.5em minus 0.4em\relax IEEE, 2021, pp. 755--760.

\bibitem{stratil2021supersingular}
P.~Stratil, S.~Hasegawa, and H.~Shizuya, ``Supersingular isogeny-based cryptography: A survey,'' \emph{Interdisciplinary information sciences}, vol.~27, no.~1, pp. 1--23, 2021.

\bibitem{costello2021case}
C.~Costello, ``The case for sike: a decade of the supersingular isogeny problem,'' \emph{Cryptology ePrint Archive}, 2021.

\bibitem{mishra2025survey}
S.~Mishra, B.~Mondal, and R.~K. Jha, ``A survey on isogeny-based cryptographic protocols,'' \emph{Wireless Networks}, vol.~31, no.~3, pp. 2993--3024, 2025.

\bibitem{malina2019towards}
L.~Malina, S.~Ricci, P.~Dzurenda, D.~Smekal, J.~Hajny, and T.~Gerlich, ``Towards practical deployment of post-quantum cryptography on constrained platforms and hardware-accelerated platforms,'' in \emph{International Conference on Information Technology and Communications Security}.\hskip 1em plus 0.5em minus 0.4em\relax Springer, 2019, pp. 109--124.

\bibitem{sood2024cryptography}
N.~Sood, ``Cryptography in post quantum computing era,'' \emph{Available at SSRN 4705470}, 2024.

\bibitem{caruso2024post}
G.~Caruso, ``Post-quantum algorithms support in trusted execution environment,'' Ph.D. dissertation, Politecnico di Torino, 2024.

\bibitem{newhouse1800migration}
W.~Newhouse, M.~Souppaya, W.~Barker, C.~Brown, P.~Kampanakis, J.~Goodman, J.~Prat, J.~Gray, M.~Ounsworth, C.~Viana \emph{et~al.}, ``Migration to post-quantum cryptography quantum readi,'' \emph{NIST SPECIAL PUBLICATION}, p. 38C, 2023.

\bibitem{ott2019identifying}
D.~Ott, C.~Peikert \emph{et~al.}, ``Identifying research challenges in post quantum cryptography migration and cryptographic agility,'' \emph{arXiv preprint arXiv:1909.07353}, 2019.

\bibitem{kampanakis2018viability}
P.~Kampanakis, P.~Panburana, E.~Daw, and D.~Van~Geest, ``The viability of post-quantum x. 509 certificates,'' \emph{Cryptology ePrint Archive}, 2018.

\bibitem{kampanakis2023vision}
P.~Kampanakis and T.~Lepoint, ``Vision paper: Do we need to change some things? open questions posed by the upcoming post-quantum migration to existing standards and deployments,'' in \emph{International Conference on Research in Security Standardisation}.\hskip 1em plus 0.5em minus 0.4em\relax Springer, 2023, pp. 78--102.

\bibitem{berbecaru2023evaluation}
D.~G. Berbecaru and A.~Lioy, ``An evaluation of x. 509 certificate revocation and related privacy issues in the web pki ecosystem,'' \emph{IEEE Access}, vol.~11, pp. 79\,156--79\,175, 2023.

\bibitem{olutimehin2025future}
A.~T. Olutimehin, S.~Joseph, A.~J. Ajayi, O.~C. Metibemu, A.~Y. Balogun, and O.~O. Olaniyi, ``Future-proofing data: Assessing the feasibility of post-quantum cryptographic algorithms to mitigate ‘harvest now, decrypt later’attacks,'' \emph{Decrypt Later’Attacks (February 17, 2025)}, 2025.

\bibitem{dowling2018cryptographic}
B.~Dowling and K.~G. Paterson, ``A cryptographic analysis of the wireguard protocol,'' in \emph{International Conference on Applied Cryptography and Network Security}.\hskip 1em plus 0.5em minus 0.4em\relax Springer, 2018, pp. 3--21.

\bibitem{baseri2024navigating}
Y.~Baseri, V.~Chouhan, and A.~Hafid, ``Navigating quantum security risks in networked environments: A comprehensive study of quantum-safe network protocols,'' \emph{Computers \& Security}, p. 103883, 2024.

\bibitem{malina2021post}
L.~Malina, P.~Dzurenda, S.~Ricci, J.~Hajny, G.~Srivastava, R.~Matulevi{\v{c}}ius, A.-A.~O. Affia, M.~Laurent, N.~H. Sultan, and Q.~Tang, ``Post-quantum era privacy protection for intelligent infrastructures,'' \emph{IEEE Access}, vol.~9, pp. 36\,038--36\,077, 2021.

\bibitem{sargiotis2024legal}
D.~Sargiotis, ``Legal and regulatory considerations in data governance,'' in \emph{Data Governance: A Guide}.\hskip 1em plus 0.5em minus 0.4em\relax Springer, 2024, pp. 445--466.

\bibitem{lonc2023feasibility}
B.~Lonc, A.~Aubry, H.~Bakhti, M.~Christofi, and H.~A. Mehrez, ``Feasibility and benchmarking of post-quantum cryptography in the cooperative its ecosystem,'' in \emph{2023 IEEE Vehicular Networking Conference (VNC)}.\hskip 1em plus 0.5em minus 0.4em\relax IEEE, 2023, pp. 215--222.

\bibitem{ali2021pragmatic}
A.~Ali, ``A pragmatic analysis of pre-and post-quantum cyber security scenarios,'' in \emph{2021 International Bhurban Conference on Applied Sciences and Technologies (IBCAST)}.\hskip 1em plus 0.5em minus 0.4em\relax IEEE, 2021, pp. 686--692.

\bibitem{pandeya2021strategy}
G.~R. Pandeya, T.~U. Daim, and A.~Marotzke, ``A strategy roadmap for post-quantum cryptography,'' \emph{Roadmapping Future: Technologies, Products and Services}, pp. 171--207, 2021.

\bibitem{alghamdi2025post}
A.~Alghamdi, ``Post-quantum cryptography implementation challenges: Security implications for critical infrastructure,'' 2025.

\bibitem{dekker2022regulating}
T.~Dekker and F.~Martin-Bariteau, ``Regulating uncertain states: A risk-based policy agenda for quantum technologies,'' 2022.

\bibitem{adeyinka2025cybersecurity}
K.~I. Adeyinka and T.~I. Adeyinka, ``Cybersecurity measures for protecting data,'' in \emph{Analyzing Privacy and Security Difficulties in Social Media: New Challenges and Solutions}.\hskip 1em plus 0.5em minus 0.4em\relax IGI Global Scientific Publishing, 2025, pp. 365--414.

\bibitem{adapa2025architecting}
V.~R.~K. Adapa, ``Architecting quantum-resistant cybersecurity: A framework for transitioning to post-quantum cryptographic systems,'' 2025.

\bibitem{fernandez2019pre}
T.~M. Fern{\'a}ndez-Caram{\'e}s, ``From pre-quantum to post-quantum iot security: A survey on quantum-resistant cryptosystems for the internet of things,'' \emph{IEEE Internet of Things Journal}, vol.~7, no.~7, pp. 6457--6480, 2019.

\bibitem{shamo2024bridging}
S.~A. Shamo, ``Bridging the quantum divide: A comprehensive analysis of nist and iso standards for post-quantum cryptography and strategies for global harmonization,'' \emph{Available at SSRN 4864519}, 2024.

\bibitem{xie2020special}
J.~Xie, K.~Basu, K.~Gaj, and U.~Guin, ``Special session: The recent advance in hardware implementation of post-quantum cryptography,'' in \emph{2020 IEEE 38th VLSI Test Symposium (VTS)}.\hskip 1em plus 0.5em minus 0.4em\relax IEEE, 2020, pp. 1--10.

\bibitem{gharavi2024post}
H.~Gharavi, J.~Granjal, and E.~Monteiro, ``Post-quantum blockchain security for the internet of things: Survey and research directions,'' \emph{IEEE Communications Surveys \& Tutorials}, vol.~26, no.~3, pp. 1748--1774, 2024.

\bibitem{algazy2024development}
K.~Algazy, K.~Sakan, A.~Khompysh, and D.~Dyusenbayev, ``Development of a new post-quantum digital signature algorithm: Syrga-1,'' \emph{Computers}, vol.~13, no.~1, p.~26, 2024.

\bibitem{alagic2019status}
G.~Alagic, G.~Alagic, J.~Alperin-Sheriff, D.~Apon, D.~Cooper, Q.~Dang, Y.-K. Liu, C.~Miller, D.~Moody, R.~Peralta \emph{et~al.}, ``Status report on the first round of the nist post-quantum cryptography standardization process,'' 2019.

\bibitem{nookala2024post}
G.~Nookala, K.~R. Gade, N.~Dulam, and S.~K.~R. Thumburu, ``Post-quantum cryptography: Preparing for a new era of data encryption,'' 2024.

\bibitem{gbajadesigning}
C.~Gbaja, ``Designing quantum-resilient data encryption protocols for securing multi-cloud architectures in critical infrastructure networks.''

\bibitem{zhang2020depth}
K.~Zhang and V.~E. Korepin, ``Depth optimization of quantum search algorithms beyond grover's algorithm,'' \emph{Physical Review A}, vol. 101, no.~3, p. 032346, 2020.

\bibitem{radanliev2024ethics}
P.~Radanliev, O.~Santos, A.~Brandon-Jones, and A.~Joinson, ``Ethics and responsible ai deployment,'' \emph{Frontiers in Artificial Intelligence}, vol.~7, p. 1377011, 2024.

\bibitem{kannwischer2022improving}
M.~J. Kannwischer, P.~Schwabe, D.~Stebila, and T.~Wiggers, ``Improving software quality in cryptography standardization projects,'' in \emph{2022 IEEE European Symposium on Security and Privacy Workshops (EuroS\&PW)}.\hskip 1em plus 0.5em minus 0.4em\relax IEEE, 2022, pp. 19--30.

\bibitem{oliva2024cybersecurity}
J.~Oliva~delMoral, A.~deMarti iOlius, G.~Vidal, P.~M. Crespo, and J.~E. Martinez, ``Cybersecurity in critical infrastructures: A post-quantum cryptography perspective,'' \emph{IEEE Internet of Things Journal}, 2024.

\bibitem{purohit2024building}
A.~Purohit, M.~Kaur, Z.~C. Seskir, M.~T. Posner, and A.~Venegas-Gomez, ``Building a quantum-ready ecosystem,'' \emph{IET Quantum Communication}, vol.~5, no.~1, pp. 1--18, 2024.

\end{thebibliography}

\end{document}